\documentclass[fleqn,usenatbib,useAMS]{mnras}

\usepackage{newtxtext,newtxmath}

\usepackage[T1]{fontenc}
\usepackage{ae,aecompl}
\usepackage{newtxtext,newtxmath}

\DeclareRobustCommand{\VAN}[3]{#2}
\let\VANthebibliography\thebibliography
\def\thebibliography{\DeclareRobustCommand{\VAN}[3]{##3}\VANthebibliography}

\usepackage{graphicx}	
\usepackage{amsmath}	
\usepackage{booktabs}
\usepackage{multirow}
\usepackage{siunitx}
\usepackage{textgreek}

\usepackage[export]{adjustbox}

\renewcommand{\epsilon}{\varepsilon}
\newcommand\smaller[2][0.85]{{\scalefont{#1}#2}}
\newcommand\dppp{\textsc{dp}\smaller[0.76]{3}}
\newcommand\dsnine{\textsc{ds}\smaller[0.76]{9}}

\title[Physical spectral index modelling]{A novel radio imaging method for physical spectral index modelling}

\author[E. Ceccotti et al.]{E.\ Ceccotti$^{1}$\thanks{E-mail: ceccotti@astro.rug.nl},
A.~R.\ Offringa$^{2,1}$,
L.~V.~E.\ Koopmans$^{1}$,
R.\ Timmerman$^{3}$,
S.~A.\ Brackenhoff$^{1}$,
B.~K.\ Gehlot$^{1}$,
\newauthor
F.~G.\ Mertens$^{4}$,
S.\ Munshi$^{1}$,
V.~N.\ Pandey$^{2,1}$, 
R.~J.\ van Weeren$^{3}$,
S.\ Yatawatta$^{2}$, 
S.\ Zaroubi$^{5,6,1}$
\\
$^{1}$Kapteyn Astronomical Institute, University of Groningen, PO Box 800, 9700 AV Groningen, The Netherlands\\ 
$^{2}$ASTRON, PO Box 2, 7990 AA Dwingeloo, The Netherlands\\
$^{3}$Leiden Observatory, Leiden University, PO Box 9513, 2300 RA Leiden, The Netherlands\\
$^{4}$LERMA, Observatoire de Paris, PSL Research University, CNRS, Sorbonne Universit\'{e}, F-75014 Paris, France\\
$^{5}$Department of Natural Sciences, The Open University of Israel, 1 University Road, PO Box 808, Ra’anana 4353701, Israel\\
$^{6}$Department of Physics, Technion, Haifa 32000, Israel
}

\date{Accepted 2023 August 11. Received 2023 August 11; in original form 2023 March 15}

\pubyear{2023}

\begin{document}
\label{firstpage}
\pagerange{\pageref{firstpage}--\pageref{lastpage}}
\maketitle

\begin{abstract}
We present a new method, called ``forced-spectrum fitting'', for physically-based spectral modelling of radio sources during deconvolution. This improves upon current common deconvolution fitting methods, which often produce inaccurate spectra. Our method uses any pre-existing spectral index map to assign spectral indices to each model component cleaned during the multi-frequency deconvolution of \textsc{wsclean}, where the pre-determined spectrum is fitted. The component magnitude is evaluated by performing a modified weighted linear least-squares fit. We test this method on a simulated LOFAR-HBA observation of the 3C\,196 QSO and a real LOFAR-HBA observation of the 4C+55.16 FRI galaxy. We compare the results from the forced-spectrum fitting with traditional joined-channel deconvolution using polynomial fitting. Because no prior spectral information was available for 4C+55.16, we demonstrate a method for extracting spectral indices in the observed frequency band using ``clustering''. The models generated by the forced-spectrum fitting are used to improve the calibration of the datasets. The final residuals are comparable to existing multi-frequency deconvolution methods, but the output model agrees with the provided spectral index map, embedding correct spectral information. While forced-spectrum fitting does not solve the determination of the spectral information itself, it enables the construction of accurate multi-frequency models that can be used for wide-band calibration and subtraction.

\end{abstract}





\section{Introduction}

Modern radio interferometers reach high spatial resolutions and have large instantaneous bandwidth, allowing the generation of sky images with high dynamic range. To achieve this, a meticulous sky model is needed to calibrate data, correcting for effects that occur along the signal path \citep[e.g.,][]{smirnov:2011}. This requires accurate modelling of all bright radio sources within a wide field of view, many of which have complex spatial and spectral structures. 

Discrete radio sources, such as supernova remnants and radio galaxies, can be modelled by shapelets, which are basis functions constituted by weighted Hermite polynomials that have an analytical Fourier transform \citep{refregier:2003,yatawatta:2011}. This property makes such functions attractive, because radio interferometers measure visibilities in the Fourier domain -- the so-called $uv$-space -- of the image brightness distribution. Therefore, shapelets are becoming increasingly more common in calibration software, such as \textsc{sagecal}\footnote{\url{https://github.com/nlesc-dirac/sagecal}} \citep{kazemi_etal:2011,yatawatta:2015,yatawatta:2016,yatawatta:2017} and \textsc{rts} \citep{mitchell_etal:2008,riding_etal:2017}, especially because it is possible to use a single basis set of functions to efficiently model extended emission in the sky \citep[e.g.,][]{gehlot_etal:2022}. However, to our knowledge, there is no direct implementation yet of shapelets modelling during the cleaning process of imaging. Such models are often produced afterwards through linear least-squares fitting of the basis functions to the final deconvolved image, using software such as \textsc{shamfi}\footnote{ \url{https://github.com/JLBLine/SHAMFI}} \citep{line_etal:2020}, \textsc{pybdsf}\footnote{\url{https://www.astron.nl/citt/pybdsf}} \citep{mohan_rafferty:2015} or \textsc{shapelet\_gui}\footnote{\url{https://github.com/SarodYatawatta/shapeletGUI}} \citep{yatawatta:2013}. 

In principle, shapelets can be used in compressive sensing techniques, as pointed out by \citet{dabbech_etal:2015}. These emerging techniques are based on convex optimisation algorithms and signal sparsity to reconstruct the sky model \citep[e.g.,][]{wiaux_etal:2009,li_etal:2011,carrillo_etal:2012,garsden_etal:2015,birdi_etal:2020,terris_etal:2022}, where isotropic undecimated wavelets \citep{starck_etal:2007} are usually used instead of shapelets. Compressive sensing is very promising in capturing the finer source details \citep[e.g.,][]{dabbech_etal:2018,dabbech_etal:2022}, but at the cost of high computational and memory requirements. 

One of the most common alternative methods -- and less computationally demanding -- is the use of \textsc{clean}-based algorithms \citep{hogbom:1974,clark:1980,schwab:1984}, specifically multi-scale (MS) deconvolution, where a source is modelled as a summation of components with different scales \citep{cornwell:2008}. As shown by \citet{offringa_smirnov:2017}, MS cleaning is particularly efficient when Gaussian components are used, because their Fourier transform is also Gaussian. The combination of Gaussian and point components (i.e., delta-functions) allows one to capture the finer structures of sky sources, but at the cost of a number of components that is usually much higher than using shapelets. \citet{line_etal:2020} show that shapelets models leave lower residuals than MS cleaning on the larger spatial scales when these methods are applied to simulated data of radio galaxies. However, results on real data applications show negligible differences because residuals are dominated by other uncertainties, like calibration errors \citep[e.g.,][]{gehlot_etal:2022}. Furthermore, current implementations of shapelets do not allow varying spectral indices across the source, unlike Gaussian and point components, where each component can have a different spectral index value. Some compressive sensing-based methods \citep[e.g.,][]{ferrari_etal:2015,abdulaziz_etal:2019} and other non \textsc{clean}-based techniques, such as maximum entropy methods \citep{bajkova_pushkarev:2011} and Bayesian inference techniques \citep{junklewitz_etal:2015}, can simultaneously model sky-brightness and spectral features, but they have only been tested on simple cases. Moreover, these do not have the advantage of separating the model into components that can easily be inverted to Fourier space. For this reason, MS deconvolution is the de facto method when accurate modelling is required at different frequencies for wide band observations. 

In this context, multi-frequency (MF) deconvolution algorithms are important, because they allow fitting polynomial functions during deconvolution to simultaneously infer spectral information and increase the deconvolution accuracy \citep{sault_wieringa:1994}. Combining MS with MF algorithms allows for accurate cleaning of resolved and diffuse sources, while simultaneously inferring spectral information \citep{rau_cornwell:2011,offringa_smirnov:2017}. However, current fitting methods generate unphysical spectral index values, especially for faint and complex (resolved) sources \citep{rau_etal:2016}. This mainly occurs because real data have systematic errors that can cause the fit to diverge, especially at the band edges, producing inaccurate spectral indices \citep[e.g.,][]{offringa_etal:2016}. For this reason, spectral information is usually extracted after the imaging process and can be inserted into calibration models afterwards.

Extracting spectral index maps from in-band observations is challenging. Inaccuracies during calibration, such as flux scale alignment \citep[e.g.,][]{shimwell_etal:LoTSS_V:2022}, or imaging, such as spectral fitting during MF deconvolution \citep[e.g.,][]{heywood_etal:2016}, can limit the reliability of in-band spectral indices, with errors often exceeding 20\% \citep[e.g.,][]{shimwell_etal:LoTSS_V:2022}. Furthermore, inaccuracy in modelling the primary beam can introduce time-dependent spectral features that are too complex to separate from real source components \citep[e.g.,][]{bhatnagar_etal:2008,tasse_etal:2013,cotton_mauch:2021}.  For these reasons, even the relatively large bandwidth ratio of 2:1 that is reached by telescopes such as LOFAR\footnote{Low-Frequency Array, \url{http://www.lofar.org}} \citep{vanhaarlem_etal:2013} and the VLA\footnote{Very Large Array, \url{https://public.nrao.edu/telescopes/vla}} \citep{perley_etal:2011}, is not always enough to properly model spectral features. Nonetheless, a few in-band spectral index analysis have been performed with satisfactory results, especially for strong sources \citep[e.g.,][]{mckean_etal:2016,arias_etal:2018,fanaroff_etal:2021,baghel_etal:2023} and source catalogues, such as GLEAM\footnote{GaLactic and Extragalactic All-sky Murchison Widefield Array, \url{https://heasarc.gsfc.nasa.gov/W3Browse/all/gleamegcat.html}} \citep{hurley-walker_etal:2017, hurley-walker_etal:2022,callingham_etal:2017} and LoTSS\footnote{LOFAR Two-metre Sky Survey, \url{https://repository.surfsara.nl/collection/lotss-dr2}} \citep{shimwell_etal:LoTSS_I:2017,shimwell_etal:LoTSS_II:2019,shimwell_etal:LoTSS_V:2022}.

The common way to extract a spectral index map is to use different instruments, where a single frequency-integrated image is considered for each wide band. In this way, any in-band uncertainty is reduced and higher signal-to-noise images can be obtained \citep[e.g.,][]{degasperin_etal:2018, digennaro_etal:2021, fanaroff_etal:2021, ignesti_etal:2022, timmerman_etal:2022a, timmerman_etal:2022b}. The downside is that the spatial resolution and sensitivity of such different telescopes should be as similar as possible. Applying a baseline cut and/or applying some smoothing kernel is always required to make sure that the same angular scales are sampled by different telescopes, which have a different array layout and operate at different frequencies \citep[e.g.,][]{vollmer_etal:2005}. This applies also to the in-band spectral index map. However, in such a case, only the change of the $uv$-coverage with frequency must be taken into account and the strength of this correction is usually lower than the one required for multi-instruments observations. This allows in principle for better capture of the finer structures in the spectral index maps, since the resolution is not excessively downgraded \citep[e.g.,][]{fanaroff_etal:2021}.

In this paper, we present a novel method, called forced-spectrum fitting, to transfer spectral information from pre-existing spectral index maps into a sky model directly during the deconvolution. This is performed with the MF algorithm of the \textsc{wsclean} imaging software \citep{offringa_etal:2014}, which cleans the image at all frequencies simultaneously, assuming only a single integrated spatial map at a reference frequency, where the clean components are defined. During this step, the forced-spectrum method assigns a power law spectral index, based on an initial spectral index map, to each of the cleaned components, rather than clean per frequency channel and derive the spectral index afterwards, since this would be affected by differences in, for example, beam, $uv$-coverage, and flagging. The initial spectral index map could be obtained from a different telescope to start with, or from in-band observations themselves, for example making a weighted average of the sky-brightness within certain regions of the source (a method that we call clustering) to reduce the effect of calibration and deconvolution errors. The data can then be calibrated with the forced-spectrum output model, the spectral index map could be adjusted based on the image residuals, and the process can be repeated, as in self-calibration. 

The proposed method promises accurate spectral index modelling without the problems related to common MF fitting, especially for extended sources. In fact, MS-MF deconvolution of these sources is less stable than deconvolving point sources, because the degrees of freedom are higher. Such degrees of freedom are reduced by the forced-spectrum method, increasing the stability of the MS-MF deconvolution.
This is particularly important for 21-cm experiments, which aim to detect the redshifted 21-cm line emitted by the neutral hydrogen during the Epoch of Reionization (EoR) and Cosmic Dawn \citep[see, e.g.,][for a review]{liu_shaw:2020}. One of the key aspects of these experiments is the separation of the foreground (Galactic and extra-galactic) emission from the redshifted 21-cm line. While the former is expected to be smooth over tens of MHz, the latter rapidly fluctuates over MHz-scales \citep[e.g.,][]{shaver_etal:1999,jelic_etal:2008}. Even with many different techniques that can remove the (residual) foreground emission \citep[e.g.,][]{parsons_backer:2009, chapman_etal:2012, bonaldi_brown:2015,mertens_etal:2018,ewall-wice_etal:2021}, spectrally accurate models are still required to improve the calibration process, where errors must be lower than $0.1\%$ to achieve sufficient dynamic range for measuring the 21-cm power spectrum during the EoR \citep{mazumder_etal:2022}.

In Sec.~\ref{sec:method} we outline the forced-spectrum method and its implementation in the MS and MF deconvolution algorithms of \textsc{wsclean}. In Sec.~\ref{sec:dataset} we introduce the simulated and real data that we use to test our method, whose results are presented in Sec.~\ref{sec:results}. Finally, in Sec.~\ref{sec:conclusions}, we discuss the results and draw conclusions.

\section{Method}\label{sec:method}

In this section, we describe a new deconvolution technique called forced-spectrum fitting, which employs a spectral index map to ensure each clean component has a specific spectral index. By using an input spectral index map with physical values, this method generates models with physical spectral information directly during the deconvolution.\footnote{Throughout this paper, we will use the term ``physical'' to refer to realistic, not extreme spectral index values that do not defy plasma physics as we currently know.} This approach overcomes the limitations of conventional fitting methods that can incorporate errors and incomplete data, resulting in clean components with unrealistic spectral indices (see further details Sec.~\ref{sec:method_fs}). 

The forced-spectrum method has been implemented in \textsc{wsclean} as an improvement of its multi-frequency (MF) deconvolution. Before describing the forced-spectrum method, we briefly illustrate how the MF deconvolution works, especially in combination with multi-scale (MS) deconvolution.

\subsection{MS-MF deconvolution in \textsc{wsclean}}\label{sec:method_wsclean}

\textsc{wsclean}\footnote{\url{https://gitlab.com/aroffringa/wsclean}} is a fast wide-field imager that uses the $w$-stacking algorithm to correct the $w$-term in wide-field radio interferometric imaging \citep{offringa_etal:2014}. Its MS and MF algorithms \citep{offringa_smirnov:2017} allow deep imaging, generating images with a high dynamic range from which spectral information can be inferred. 

The MS method generates a source model that is a summation of point components and basis functions -- such as tapered quadratic or circular Gaussian functions -- of different sizes \citep[e.g.,][]{cornwell:2008}. This reduces negative artefacts around bright resolved sources and allows for a better recovery of extended sources \citep{rich_etal:2008}. Gaussian functions are often preferred because they have an analytically defined Fourier transform.

The MF deconvolution splits a wide frequency band into output channels (i.e., subsets of narrower bandwidth), which are imaged separately. These images are combined in a frequency-integrated continuum image, which has a higher dynamic range, where the peak-finding is performed during the cleaning process. The brightness of a cleaned component is measured in the output images at each frequency channel at the location of the identified peak, thereby allowing the components to model spectral variations. In \textsc{wsclean}, this approach is called joined-channel deconvolution. 

The quality of the integrated image can be improved by using the MF weighting, which grids the weights of all output channels in a single grid to ensure the desired weighting (e.g., uniform) for the integrated image. However, this approach may cause the weights of individual frequencies to deviate, leading to artificial spectral structures due to potentially large deviations in the synthesised beam over frequency.\footnote{\url{https://wsclean.readthedocs.io/en/latest/mf_weighting.html}} Therefore, in this study, we do not use MF weighting and instead grid the weights of each output channel on separate grids, which allows us to obtain the desired weighting (e.g., uniform) for each individual image.

In MF deconvolution, a polynomial function can be fitted to the measurements to enforce spectral smoothness. In this case, the brightness subtracted from each individual output image is given by the fitted function. \textsc{wsclean} supports two fitting functions, an ordinary polynomial and a logarithmic polynomial, both requiring as extra parameter the number of terms $n$ (with $n\in\mathbb{Z}^+$) at which the Taylor expansion is truncated, giving a $(n-1)$th-order polynomial. The ordinary polynomial at frequency $\nu$ is given by
\begin{equation}\label{eq:pol}
    S(\nu) = \sum_{i=0}^{n-1} p_i\left ( \frac{\nu}{\nu_0} -1\right)^{i}\, ,
\end{equation}
where $S$ is the flux density and $\nu_0$ is the reference frequency at which coefficients $p_i$ are evaluated.\footnote{Here we define $0^0=1$, as \textsc{wsclean} and \textsc{python} do.} The $-1$ within brackets is used to make sure that $p_0$ gives the flux density value $S(\nu_0)$ at the reference frequency. The logarithmic polynomial of $(n-1)$th-order is given by
\begin{equation}\label{eq:logpol}
    S(\nu) = \prod_{i=0}^{n-1} 10 ^{c_i\log_{10}^i\left(\nu/\nu_0\right)}\, ,
\end{equation}
where $c_i$ are the coefficients. Note that $c_0=\log_{10} S(\nu_0)$ and $c_1=\alpha$, where $\alpha$ is the spectral index. When $n=2$, Eq.~\eqref{eq:logpol} can be re-written as a power law:
\begin{equation}\label{eq:powerlaw}
    S(\nu) = S(\nu_0)\left(\frac{\nu}{\nu_0}\right)^\alpha\,.
\end{equation}
Whatever function is used, the fitting parameters are used to estimate the flux density at the central frequency $\nu$ of each output channel and the smooth components are added to the model. During deconvolution, ordinary polynomial functions are generally preferred over logarithmic polynomials, where negative artefacts, which could be picked up by the cleaning, lead to high values at the edge channels, causing instability. 

When the fitting is turned on during the MF deconvolution, \textsc{wsclean} can generate a model catalogue of all the clean components. The catalogue consists of information such as sky coordinates, flux density, spectral shapes, etc.\ for every clean component (see Appendix~\ref{app:modeltext} for an example). This means that such a model is not limited by the image pixel scale, thereby extended components do not need to be pixelized and very high resolution models can be obtained with fewer components and hence a smaller data volume. The model catalogue can then be directly used for calibration in software such as \dppp\footnote{\url{https://dp3.readthedocs.io}} \citep[Default Pre-Processing Pipeline;][]{vandiepen_etal:2018} or can be rendered to model images with the preferred size and resolution. 

\subsection{Forced-spectrum fitting}\label{sec:method_fs}

Whereas the spectral fitting in the MF-MS deconvolution is very useful to infer spectral information, it often generates clean components with unphysical and inaccurate spectral indices. This can happen either because the frequency bandwidth is too small or because spectral smoothness is forced on data that are non-smooth by some systematic, such as beam sidelobes. Fitting errors propagate through the cleaning process, are absorbed into adjacent components, and finally lead to incorrect spectral indices. Therefore, analysing the spectral information produced in this manner is not recommended.

Forced-spectrum fitting solves this problem. It allows generating models with accurate and physically-motivated spectral information directly during the deconvolution. This method uses a pre-existing spectral index map to force spectral indices of each clean component during the deconvolution, fitting a first-order ($n=2$) logarithmic polynomial function described by Eq.~\eqref{eq:logpol}. Since $c_1$ (and optionally higher terms) are fully constrained by the input spectral index map, only $c_0$ must be fitted. This regulates the fitting and makes it more stable and faster, reducing the degrees of freedom.

To fit for the most probable value of $c_0$, a (weighted) linear least-squares fit should be performed. When we define a helper function $f(\nu) = \left(\nu / \nu_0\right)^{c_1}$ and $S_0 = S(\nu_0) = 10^{c_0}$, the most probable value $\hat{S}_0$ is given by
\begin{equation}\label{eq:least-squares-forced-spectrum-fitting}
  \hat{S}_0 = \frac{\sum_i w_i s_i f(\nu_i)}{\sum_i w_i f^2(\nu_i)} \, .
\end{equation}
Here, the sum is over the frequency direction of one pixel, with frequencies $\nu_i$ and weight $w_i = 1 / \sigma_i^2$, the inverse variance of measurement $s_i$. However, the use of this function inside joined-channel deconvolution leads to instability. The reason for this is that, during peak-finding, the peak is selected that maximises $\big|\sum_i w_i s_i \big|$. Because Eq.~\ref{eq:least-squares-forced-spectrum-fitting} is not weighted by $f(\nu_i)$, it can happen that a peak is selected for which the weighted squared difference $w_i \left(s_i - c_0 f(\nu_i)\right)^2$ is already minimised, causing the fitted value $\hat{S}_0$ to be (almost) zero. As a consequence, the residual image remains (nearly) unchanged after the iteration and the same peak is found in the next iteration. This causes an infinite loop, ultimately resulting in the deconvolution process becoming stuck. We observe that this situation is triggered in almost any reasonably deep deconvolution run (several thousand iterations) when implementing forced-spectrum fitting as in Eq.~\eqref{eq:least-squares-forced-spectrum-fitting}. Therefore, this is a problem that is faced in any application in practice.

One solution to this problem is to simply not weigh the fitting function by $f(\nu_i)$, and instead fit $\hat{S}_0$ using
\begin{equation} \label{eq:stable-forced-spectrum-fitting}
  \hat{S}_0 = \frac{\sum_i w_i s_i}{\sum_i w_i f(\nu_i)}\,.
\end{equation}
Effectively, this performs the fit using a modified weight $w'_i = w_i f(\nu_i)$. Because peak-finding now selects the pixel that maximises the absolute numerator in this equation, each deconvolution iteration selects the pixel that maximises $\left|\hat{S}_0\right|$, thereby guaranteeing that progress is made. The downside of this choice is that the fitted spectrum is not optimally weighted. For example, when deconvolving a steep-spectrum source ($c_1 < 0$), high frequencies will be over-weighted, causing its error (caused by noise, sidelobes and calibration artefacts) to have a larger effect on $\hat{S}_0$. This effect would be strongest when using a large bandwidth with steep or strongly inverted sources. The effect that sidelobes from other sources have is mitigated by the iterative nature of \textsc{clean}-based deconvolution, which involves ``revisiting'' a pixel in later iterations. This does of course not hold for noise and calibration artefacts. As a consequence, the use of forced-spectrum fitting requires a higher sensitivity and calibration quality compared to performing the same fit on an image cube outside of deconvolution. This is of course not so surprising, given that spectral deconvolution has to solve a more complex problem. We will analyse the accuracy of deconvolution in Sec.~\ref{sec:results}.

To also force higher-order terms, such as spectral curvature, function $f$ in Eq.~\eqref{eq:stable-forced-spectrum-fitting} can be changed to have the form of Eq.~\eqref{eq:logpol} divided by $10^{c_0}$, i.e.,
\begin{equation}
  f(\nu_i) = \prod_{i=1}^{n-1} 10 ^{c_i\log_{10}^i\left(\nu/\nu_0\right)}\, ,
\end{equation}
where the product is over the number of terms, excluding the first term $c_0$. This is a straightforward extension that will be investigated in a future paper.

Any kind of spectral index map can be used as input of the forced-spectrum method: it can be extracted either from in-band data from the same telescope, or by combining different bands from different instruments. Moreover, smoothing can be applied to increase the robustness of the spectral index estimates. Each map pixel should have a spectral index value. This means that pixels from empty regions of the sky need a spectral index value. Because such regions will not be cleaned (or at a very low level), an arbitrary value can be assigned. For our testing purpose in Sec.~\ref{sec:results}, we decide to assign $\alpha=-0.7$, i.e., the typical value for radio sources emitting by synchrotron radiation, to areas where no spectral index value has been extracted, because we still want to assign realistic prior values to any faint source that may not have been previously cleaned. Such assumption is not strict and different observations may require different values. Spectral information are then transferred from the input map to the output models, where each component in the model catalogue will have the spectral index value of the map pixel where its coordinates lay. Such a physical model can be used for self-calibration (see Sec.~\ref{sec:4C55.16res}) or even to calibrate different datasets with accurate spectral information.  

\subsubsection{Behaviour of forced-spectrum for overlapping components}\label{sec:overlap}

\begin{figure}
    \centering
    \includegraphics[width=0.85\columnwidth]{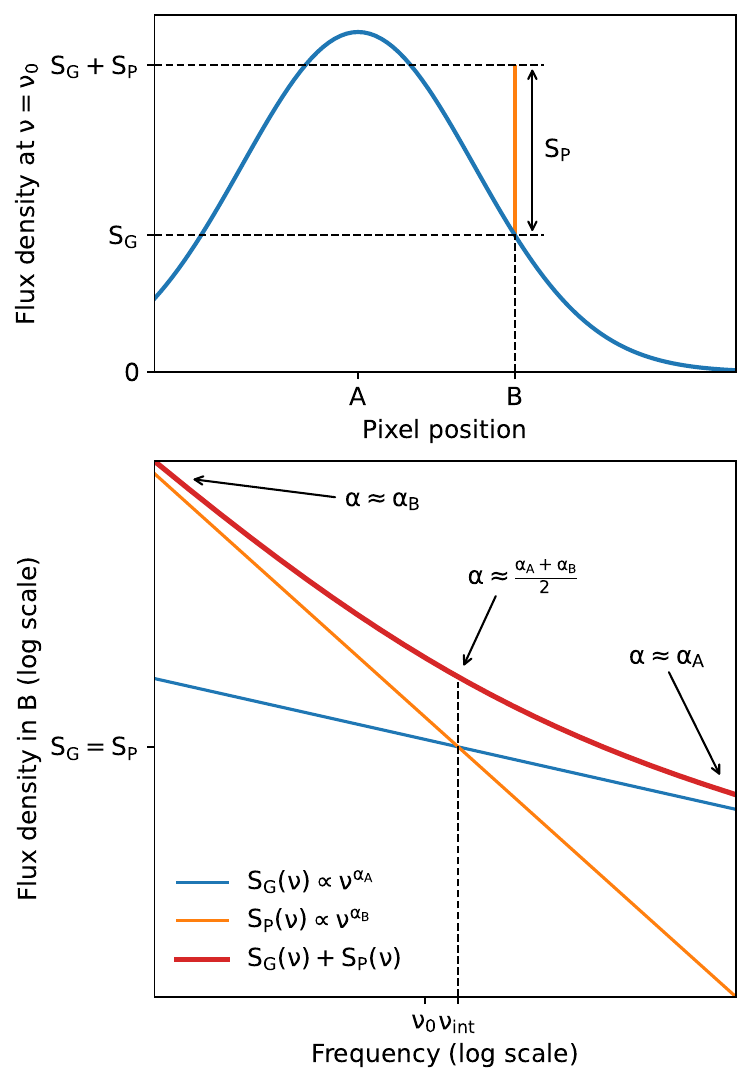}
    \caption{Example case of the overlap of a Gaussian (blue line), with the peak in A and spectral index $\alpha_\text{A}$, and a point component (orange line), located in $B$ with spectral index $\alpha_\text{B}$. The top panel shows such overlap on the image space at a reference frequency $\nu_0$, where the flux densities are $S_\text{G}$ and $S_\text{P}$ for the Gaussian and the point source, respectively. The bottom panel shows the spectra of the components in the overlapping position B, in a log-log space. The sum of the two components is plotted with the red line, which asymptotically tends to $S_\text{P}$ at low frequencies, with a slope $\alpha\approx\alpha_\text{B}$, and to $S_\text{G}$ at high frequencies, with a slope $\alpha\approx\alpha_\text{A}$. Where the single component spectra intersect, i.e., in $\nu_\text{int}$, the slope of the sum is $\alpha\approx(\alpha_\text{A} + \alpha_\text{B})/2$.}
    \label{fig:si_overlapping}
\end{figure}

In the forced-spectrum method, the spectral index of a component is assigned based on its central position. When MS deconvolution is used, extended components, such as Gaussians, will have a constant spectrum across their entire shape, determined from their central position. As a consequence, the resulting spectral index of the model may not exactly follow the input spectral index map, and particular artefacts may arise when components overlap. An example case is where one component at position A receives a spectral index $\alpha_\text{A}$ and another nearby component at position B receives a different spectral index $\alpha_\text{B}$. If the shapes of the two components overlap, the model at that location becomes a (weighted) combination of the two spectral indices at their centres.

The result of such overlap is shown in Fig.~\ref{fig:si_overlapping}, where we consider a Gaussian, with its peak at position A and spectral index $\alpha_\text{A}$, and a point component at position B with spectral index $\alpha_\text{B}$. This means that the forced-spectrum method assigns $\alpha_\text{B}$ to the point component, but the spectral index in B extracted by fitting a power law to the model images at different frequencies would be a mix of $\alpha_\text{A}$ and $\alpha_\text{B}$. The total flux density resulting in B is the sum of the contributions of the Gaussian in that pixel, $S_\text{G}$, and of the point component, $S_\text{P}$. As shown in the bottom panel of Fig.~\ref{fig:si_overlapping}, the sum of two power laws with different spectral indices is not a power law anymore. In Appendix~\ref{app:overlapping}, we derive its expected slope $\alpha$, which is a function of frequency that, when $\alpha_\text{A}>\alpha_\text{B}$, is bound in the range $\alpha_\text{A}\gtrsim \alpha \gtrsim \alpha_\text{B}$.

When component overlap occurs, the resulting spectral indices in the pixel-based output model images would not agree with the spectral index map used as input in the forced-spectrum method. If the cleaning is performed using only point components, for example, using the Cotton-Schwab algorithm, this issue is not present and the spectral indices resulting from output models exactly match the input map. As we will show in Sec.~\ref{sec:results}, the artefacts related to overlapping components are minor compared to the advantages that the combination of forced-spectrum and MS deconvolution provides. In any case, it is important to keep this phenomena in mind to fully understand our results.

\section{Datasets}\label{sec:dataset}

In this section, we discuss the datasets used to test the forced-spectrum method and compare it with traditional polynomial-based fitting. We use a simulated LOFAR High-Band Antenna (HBA) observation of the 3C\,196 quasar (QSO) and a real observation of the FRI radio galaxy 4C+55.16.

LOFAR \citep{vanhaarlem_etal:2013} is a low-frequency radio interferometer located in the north of the Netherlands and across Europe. It is constituted by two types of antennas: Low-Band Antennas (LBA), which operate from 10 to 90\,MHz, and High-Band Antennas (HBA), which operate from 110 to 240\,MHz. These two antennas types are grouped into stations. Currently, there are 24 core stations (CS; maximum baseline $\sim4\,\text{km}$) near the array centre, 14 remote stations (RS; maximum baseline $\sim120\,\text{km}$) across the Netherlands, and 14 international stations (IS; maximum baseline $\sim2000\,\text{km}$) throughout Europe. The international stations allow for very long baseline interferometry (VLBI) observations. Each CS consists of two HBA sub-stations (HBA0 and HBA1) that can be used as independent stations, increasing the number of short baselines and consequently providing a better $uv$-coverage. Additionally, all of the core stations can be combined into a phased-up super-station (ST) that has a much higher sensitivity and a narrower field of view than individual CS. This helps the calibration of international stations and improves the self-calibration, as well as reducing the data volume by about 80\% \citep{morabito_etal:2022}. 

\begin{table}
\caption{Observational details of the datasets.}
\label{tab:obs-details}
\begin{tabular}{lcc}
\toprule
\multirow{2}*{Parameter} & \multicolumn{2}{c}{Value}\\
\cmidrule(lr){2-3}
 & $\text{3C\,196}^\text{a}$ & $\text{4C+55.16}$ \\
\midrule
Telescope & LOFAR HBA & LOFAR HBA \\
Project code & LC14\_020 & LC14\_019 \\
Antenna configuration & HBA Dual Inner & HBA Dual Inner \\
Number of stations & $75$ ($\text{NL} + \text{IS}^\text{b}$) & $28$ ($\text{ST} + \text{RS} + \text{IS}^\text{b}$) \\
Obs.\ start time (UTC) & 2020 Oct 12; 20:30 & 2020 Nov 9; 23:41 \\
Phase centre (J2000): & {} & {} \\
{ }{ } Right Ascension & $08^\text{h}13^\text{m}36\rlap{.}^\text{s}07$ & $08^\text{h}34^\text{m}54\rlap{.}^\text{s}90$ \\
{ }{ } Declination & $+48^\circ13'02\rlap{.}''58$ & $+55^\circ34'21\rlap{.}''07$ \\
Duration of observation & 8\,h & 8\,h \\
Frequency range & 110--190\,MHz & 120--165\,MHz \\
Time resolution$^\text{c}$ & 20\,s & 16\,s \\
Frequency resolution$^\text{c}$ & 195\,kHz & 0.5\,MHz \\
\bottomrule
\end{tabular}
\vspace{1ex}

\hspace{1ex}{\footnotesize $^\text{a}$ Measurement set into which we predicted 3C\,196 simulations. \par}
\hspace{1ex}{\footnotesize $^\text{b}$ International station PL611 did not participate in this observation. \par}
\hspace{1ex}{\footnotesize $^\text{c}$ After flagging and averaging. \par}
\end{table}

Both simulations and observations used in this work have been obtained using international stations, reaching sub-arcsecond resolution. The observational details of the two datasets are reported in Table~\ref{tab:obs-details}. All data processing and analysis is performed on the Dawn high-performance computing cluster \citep{pandey_etal:2020} at the University of Groningen.

\subsection{Simulations of 3C 196}\label{sec:3C196sim}

\begin{figure*}
    \centering
    \includegraphics[width=1\textwidth]{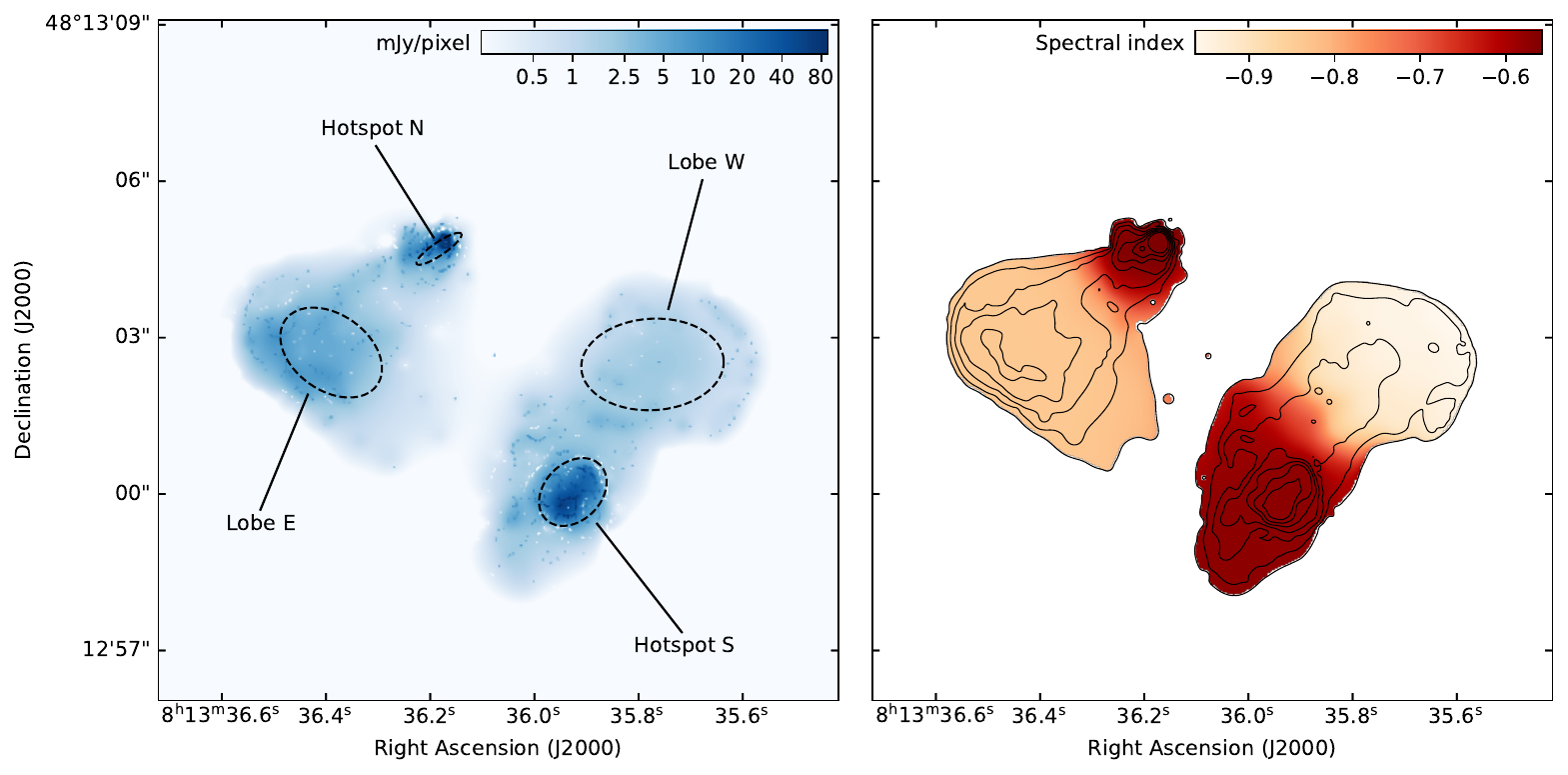}
    \caption{Rendered image of the 3C\,196 simulated model at 143\,MHz (left) and the ground-truth spectral index map (right) as obtained from the smoothed images. The Gaussians of the simple 4-component model are plotted as dashed ellipses in the left image, with sizes and names as reported in Table~\ref{tab:3C196spix}. The contours in the right figure indicate the model image after the deconvolution with the Gaussian kernel, and they increment in a geometrical progression of 2, starting at $0.40\,\text{mJy\,beam}^{-1}$. The colour scale of the spectral index map spans in the range $-0.96\le\alpha\le-0.56$.}
    \label{fig:3C196_sim}
\end{figure*}

\begin{table}
\caption{Position (RA and Dec in J2000), spectral index $\alpha$ at 150\,MHz, and shape (major and minor axes, and position angle) of the Gaussians of the 4-component model. The position angle (P.A.) is defined relative to the north of the image and not relative to the {\lq\lq{true}\rq\rq} north.}
\label{tab:3C196spix}
\resizebox{1\columnwidth}{!}{\tabcolsep=0.11cm
\begin{tabular}{lcccccc}
\toprule
Component & RA & Dec & $\alpha$ & Major axis & Minor axis & P.A. \\
 & & &  & (arcsec) & (arcsec) & (deg) \\
\midrule
Hotspot S & $08^\text{h}13^\text{m}35\rlap{.}^\text{s}925$ & $48^\circ13'00\rlap{.}''061$ & $-0.572$ & $1.476$ & $1.091$ & $135.27$ \\
Lobe W & $08^\text{h}13^\text{m}35\rlap{.}^\text{s}772$ & $48^\circ13'02\rlap{.}''507$ & $-0.973$ & $2.738$ & $1.754$ & $93.30$ \\
Hotspot N & $08^\text{h}13^\text{m}36\rlap{.}^\text{s}182$ & $48^\circ13'04\rlap{.}''725$ & $-0.557$ & $1.034$ & $0.248$ & $124.03$ \\
Lobe E & $08^\text{h}13^\text{m}36\rlap{.}^\text{s}389$ & $48^\circ13'02\rlap{.}''735$ & $-0.840$ & $1.456$ & $2.159$ & $324.58$ \\
\bottomrule
\end{tabular}}
\end{table}

To create complex and realistic simulated data, we take a high-resolution, bandwidth-integrated LOFAR model of the 3C\,196 QSO \citep[RA $08^\text{h}13^\text{m}36\rlap{.}^\text{s}07$, Dec $+48^\circ13'02\rlap{.}''58$ in J2000;][]{3C:spinrad_etal:1985,paris_etal:2014} to provide the spatial information, and apply different spectral indices to different parts of the image. The spectral information is taken from a simple, older 4-component model of the source, consisting of Gaussians whose position and shape are reported in Table~\ref{tab:3C196spix} together with the associated spectral index at a reference frequency of 150\,MHz. Despite the morphological simplicity of this model, its spectral indices are consistent with the expected synchrotron emission \citep{lonsdale_morison:1980}. The source has been divided into two hotspots, northern (N) and southern (S), and two lobes, eastern (E) and western (W), as shown with the dashed contours in Fig.~\ref{fig:3C196_sim}. The high-resolution model was obtained from a joined deconvolution of LOFAR-VLBI HBA-mid (116--171\,MHz) and HBA-high (230--244\,MHz) observations taken in 2016, using MS-MF deconvolution. The model consists of 2812 components (1369 point components and 1444 Gaussians), each one with an unphysical spectral index because of the third-order ordinary polynomial fitting performed during the MF deconvolution. For this reason, we replace the spectral indices of the complex model with those from the simple model, by linear interpolation of the spectral indices between the centres of the four Gaussian components.

We then render the composite model at two different frequencies, using a pixel scale of $30 \times 30\,\text{mas}$, to extract a pixel-by-pixel spectral index map, using Eq.~\eqref{eq:powerlaw}. During this operation we only consider pixels brighter than $0.40\,\text{mJy\,beam}^{-1}$. Since point components generate pixel-scale structures in the spectral index map, we convolve the rendered model images by a circular Gaussian filter with $\sigma=5\,\text{pixels}$ to obtain a smoother map: the resulting spectral indices span a range between $\alpha_{\text{min}}=-0.96$ and $\alpha_{\text{max}}=-0.56$. The result is a realistic, complex simulated ground-truth model, shown in Fig.~\ref{fig:3C196_sim}, which is then used to test our forced-spectrum fitting approach. A gradient between the four regions is visible in the pixel-based spectral index map (right panel), due to overlapping Gaussians, as explained in Sec.~\ref{sec:overlap}. The ground-truth model of 3C\,196 has been generated in the same way that the forced-spectrum method works, namely assigning spectral indices to each component according to its central position.

From the (non-smoothed) ground-truth model we generate visibilities. We use a realistic $uv$-coverage by predicting the model catalogue into a LOFAR measurement set, phase-rotated to the position of 3C\,196. It is an 8\,h HBA observation covering 110--190\,MHz. All core, remote, and international stations are included to achieve the resolution required to resolve the fine structure of 3C\,196. The time and frequency resolutions are 20\,s and 195\,kHz, respectively. The prediction is performed by \dppp. Any flags present in the original measurement set have been removed to ensure the best $uv$-coverage possible. 

Finally, we add white noise to each visibility by using the radiometer equation \citep[e.g.,][]{thompson_etal:2017}:
\begin{equation}\label{eq:radiometer_equation}
\sigma_\text{vis} = \frac{\text{SEFD}}{\sqrt{2 \Delta t \Delta \nu}}\,
\end{equation}
where SEFD is the system equivalent flux density, $\Delta t = 20\,\text{s}$ is the integration time of a single visibility and $\Delta \nu = 195\,\text{kHz}$ is the frequency channel width. For the sake of simplicity, we assign to each visibility an average SEFD given by a baseline constituted by only core stations, i.e., $\text{SEFD} = 3.3\,\text{kJy}$ \citep{vanhaarlem_etal:2013}. This value has been increased by 10\% to take into account projection effects that can reduce the effective collective area of the antennas when observing away from the zenith. We then find a per-visibility noise level of $\sigma_\text{vis}=1.2\,\text{Jy}$. The results of applying the forced-spectrum method to the simulated dataset of 3C\,196 will be discussed in Sec.~\ref{sec:3C196sim_res}.

\subsection{4C+55.16 observations}\label{sec:4C55.16}

\begin{figure*}
    \centering
    \includegraphics[width=1\textwidth]{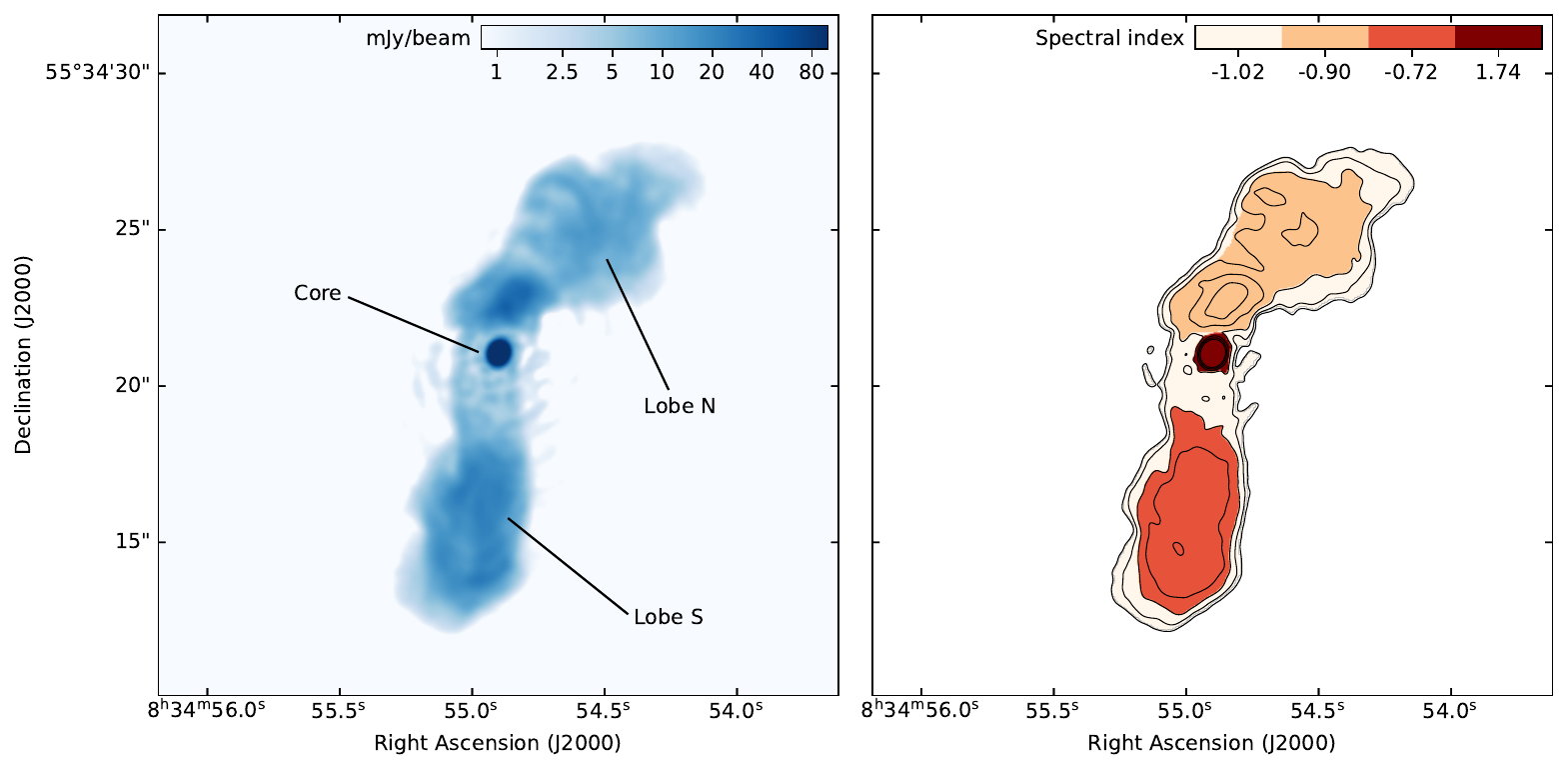}
    \caption{Integrated deconvolved image of 4C+55.16 at 143\,MHz (left) and forced-spectrum input spectral index map (right). The image noise is $\sigma=80\,\text{\textmu Jy\,beam}^{-1}$. Contours in the right figure are drawn from the integrated image and increment with a factor of 2, starting at $20\sigma$. Spectral index values are extracted from a weighted average of the source brightness within four regions: $\alpha=1.74$ for the core, $\alpha=-0.72$ for the lobe S, $\alpha=-0.90$ for the lobe N, and $\alpha=-1.02$ for the surrounding low brightness emission.}
    \label{fig:4C55_image}
\end{figure*}

In addition to the simulated observation, we use the FRI radio galaxy 4C+55.16 \citep[RA $08^\text{h}34^\text{m}54\rlap{.}^\text{s}90$, Dec $+55^\circ34'21\rlap{.}''07$ in J2000;][]{pilkington_scott:1965,charlot_etal:2020}, because it has both diffuse and compact emission regions in its lobes and core. We use an 8\,h LOFAR-VLBI HBA observation, that have been processed and calibrated by \citet{timmerman_etal:2022a}. The dataset covers 120--165\,MHz, with 16\,s and 0.5\,MHz of time and frequency resolutions, respectively. All core stations have been phased-up into a virtual super-station during the calibration of international stations. This operation reduced the field of view of the array from ${\sim}2^\circ$ to ${\sim}1.5\,\text{arcmin}$, which is sufficient because the source is only ${\sim} 18\,\text{arcsec}$ in size.

4C+55.16 has a total flux density of 8.7\,Jy at 143\,MHz, dominated by the diffuse emission from its lobes that accounts for approximately 60\% of the the total emission, while the remaining 40\% is concentrated in the core. The core consists of two components with an angular separation of 160\,mas, both with inverted spectra ($\alpha>0$) in the HBA frequency range, but with different spectral indices \citep{whyborn_etal:1985}. This, combined with a steep synchrotron emission from the lobes, produces an integrated spectral index $\alpha=-0.02$ between 143\,MHz and 1.4\,GHz -- as observed by the VLA. The two core components are not completely resolved by LOFAR, but the angular separation is comparable to the angular resolution at the higher frequencies of the observed band. This makes the phase calibration challenging and less accurate, generating imaging artefacts especially when uniform weighting is used. Strong residual sidelobes are observed when MF weighting is not used. To solve this issue, we performed several self-calibration iterations using models from the MS-MF deconvolution of \textsc{wsclean}, fitting a third-order ordinary polynomial function through the output channels. We also normalise the total flux density to the known value of 8.7\,Jy at 143\,MHz using a power law with $\alpha=-0.02$. Images are made with a pixel scale of $25\times25\,\text{mas}$, using uniform weighting without MF weighting, resulting in an integrated synthesised beam of $254.7 \times183.6\,\text{mas}$ when 92 output channels are used. We refer the reader to Appendix~\ref{app:self-cal_4c55} for more details about the self-calibration step.

The next step is to extract a spectral index map that we can use in the forced-spectrum method. We make images to extract the in-band spectral index map for the forced-spectrum method. We use the same \textsc{wsclean} settings as before, except we do not use any kind of fitting because we want the 92 channels to be completely independent. Using fewer output channels can speed up the fitting process and lower the noise per channel. However, we are currently not limited by computational power and the dynamic range of each image is high enough ($> 3000$). Therefore, we can image each of the 92 narrow channels independently. This minimises any bias that would occur during the frequency averaging due to non-linear spectral curvature over the averaged channel width. In addition, we apply a circular Gaussian taper with a full width at half-maximum (FWHM) of 250\,mas, to make sure that the two core components are not resolved, and we set a circular beam with a size of 300\,mas for restoring the clean components at the end of the deconvolution, to have the same resolution over the full bandwidth. Decreasing the resolution with the Gaussian taper increases the image noise, which goes from $67\,\text{\textmu Jy\,beam}^{-1}$ after the last self-calibration iteration (see Appendix~\ref{app:self-cal_4c55}) to $80\,\text{\textmu Jy\,beam}^{-1}$ for the obtained integrated image at 143\,MHz, shown in the left panel of Fig.~\ref{fig:4C55_image}. 

As in Sec.~\ref{sec:3C196sim}, we consider only the brightest emission of the source for the spectral index extraction, removing all the pixels less bright than $20\sigma$. Unfortunately, the $uv$-coverage is reduced when imaging the 92 channels of only 0.5\,MHz bandwidth, which causes strong sidelobes. While these sidelobes are not a problem for the self-calibration, because we can set a high cleaning thresholds to not include them in the cleaned model, they strongly affect the pixel-based in-band spectral indices of low surface brightness regions. We thus develop a clustering method to generate in-band spectral index maps. This is based on dividing the source into regions (or clusters) where the emission spectrum is assumed to have the same slope, and assigning the same spectral index to each pixel within a certain region. For 4C+55.16, we divide the source into three regions: the core, the northern (N) and the southern (S) lobes. These have been initially identified by the pixels with brightness higher than $80\sigma$ in the integrated image of Fig.~\ref{fig:4C55_image} and then shaped with small adjustments made by hand. In this case, we have performed these adjustments to remove a visible sidelobe artefact from the spectral index map area. These changes do not make a significant difference in the results, because the intensity of the image is low at the place of these changes. As a general procedure, it is probably preferred to not make such manual changes, as they may lead to a bias towards what one expects. A fourth region is added to represent the surrounding low brightness emission between $20\sigma$ and $80\sigma$. 

We then evaluate the weighted mean of the brightness $I$ within each region and per frequency channel, weighting each value by itself, i.e.,
\begin{equation}
    \overline{I}_\text{reg}(\nu) = \frac{\sum_i I_i(\nu)^2}{\sum_i I_i(\nu)}\, ,
\end{equation}
where the sum is over the number of pixels of the given region. This kind of average combines more information and reduces the contribution from the fainter components that are most affected by deconvolution artefacts. The spectral indices for each region are finally calculated from the non-linear least-squares fitting of Eq.~\eqref{eq:powerlaw}, where both $S(\nu_0)$ and $\alpha$ are evaluated at $\nu_0=150\,\text{MHz}$. In our case we can directly use the brightness $I$ (units of $\text{Jy\,beam}^{-1}$) instead of the flux density $S$ (units of Jy) because the beam size is the same for $\nu$ and $\nu_0$. The resulting spectral index map is shown in the right panel of Fig.~\ref{fig:4C55_image}, where $\alpha=1.74$ for the core, $\alpha=-0.72$ for the lobe S, $\alpha=-0.90$ for the lobe N, and $\alpha=-1.02$ for the surrounding low brightness emission. This is the map that we use as input of the forced-spectrum method.

\section{Results}\label{sec:results}

In this section, we test the forced-spectrum method on the simulated 3C\,196 and the observed 4C+55.16 datasets described in Sec.~\ref{sec:dataset}.

\subsection{Results from simulated 3C 196 observation}\label{sec:3C196sim_res}

\begin{figure*}
    \centering
    \includegraphics[width=1\textwidth]{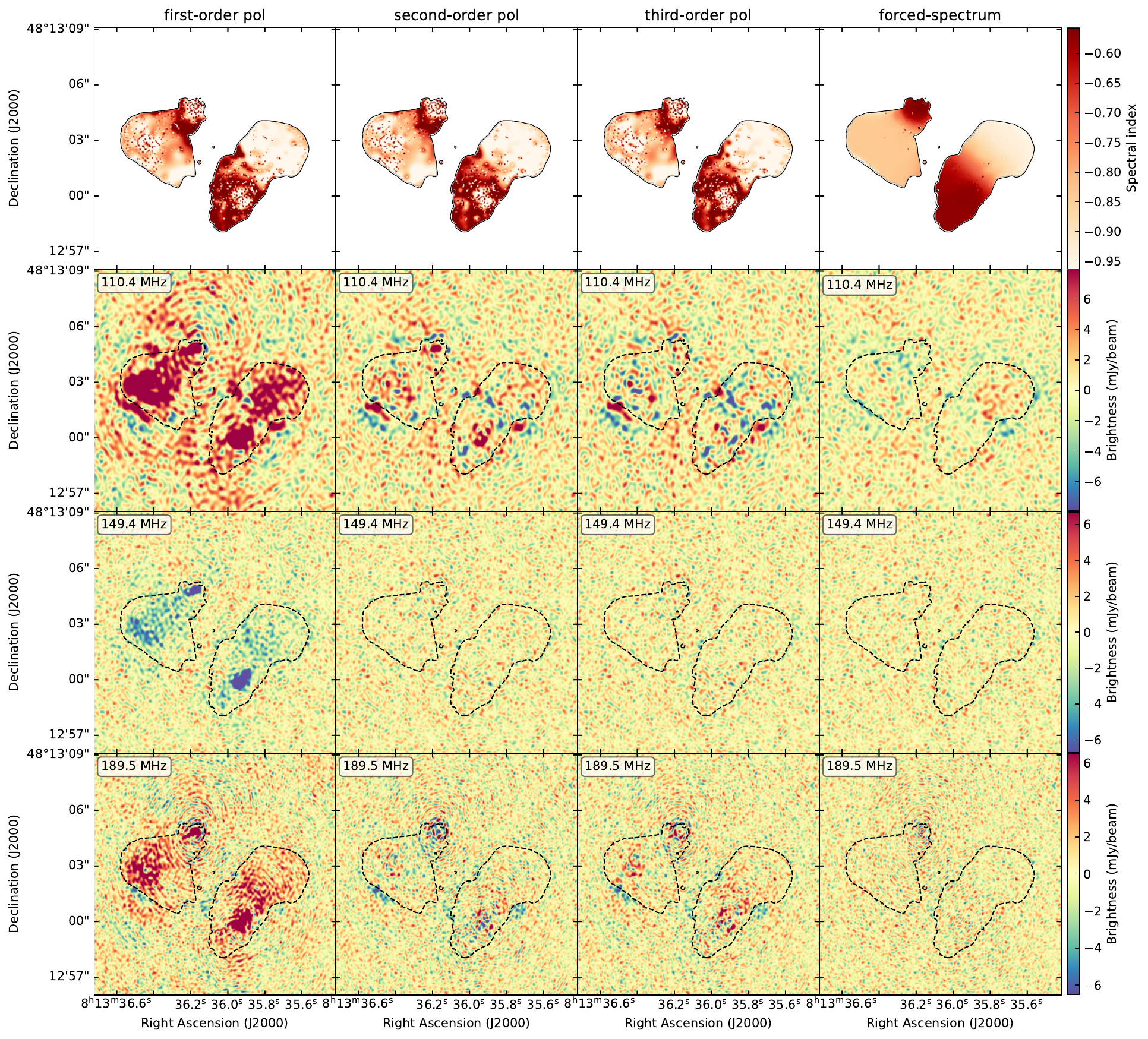}
    \caption{Results from deconvolution with different fitting methods of simulated data of 3C\,196. From left to right: results from first, second, and third-order ordinary polynomial fitting, and forced-spectrum fitting. From top to bottom: spectral index maps from output models, and residual images of the lower, middle, and higher channels, centred at 110.4, 149.4, and 189.5\,MHz, respectively. The colour scale of spectral index maps is based on the smoothed ground-truth values to make the comparison easier; however, values from ordinary polynomial fits can be outside that range, as shown in Fig.~\ref{fig:3C196_spix_distr} and in Table~\ref{tab:3C196spix_out}. On the other hand, spectral indices obtained from the forced-spectrum method are similar to the expected values, with only few outlier pixels. The dashed contours superimposed on the residual images indicate the $0.40\,\text{mJy\,beam}^{-1}$ contour level of Fig.~\ref{fig:3C196_sim}.}
    \label{fig:3C196_rescomp}
\end{figure*}

\begin{figure*}
    \centering
    \includegraphics[width=1\textwidth]{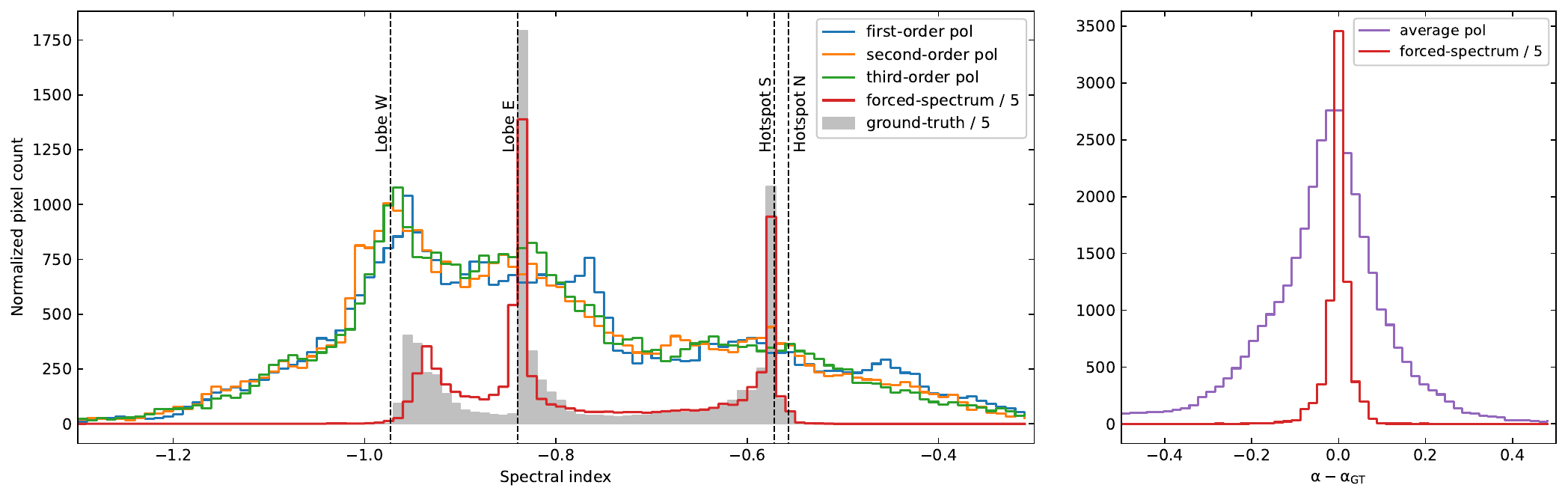}
    \caption{Pixel distributions of the spectral index maps shown in Fig.~\ref{fig:3C196_rescomp} (left) and of their difference with the non-smoothed ground-truth ($\alpha_\text{GT}$) map (right). First (blue), second (orange), and third-order (green) ordinary polynomial fitting distributions are plotted in comparison with the forced-spectrum (red) and the ground-truth (grey area) ones. In the right panel all the polynomial fitting histograms have been averaged into a single distribution (purple). Histograms are obtained by binning spectral indices into bins of 0.01 and 0.02 wide in the left and right panel, respectively. Forced-spectrum and ground-truth values have been divided by 5 to make the comparison with polynomial distributions easier. Vertical black dashed lines mark the spectral indices of the four components of 3C\,196, as reported in Table~\ref{tab:3C196spix}.}
    \label{fig:3C196_spix_distr}
\end{figure*}

To be sure that our forced-spectrum method really generates models with physical spectral information, we start testing it with the simulated 3C\,196 dataset for which we have constructed a ground-truth spectral index map. We make uniform-weighted images of the data with the same pixel scale and size of the ground-truth spectral index map. We take advantage of both MS and MF deconvolution, splitting the full bandwidth into 80 channels, each ${\sim}1\,\text{MHz}$, but turning off the MF weighting, similar to Sec.~\ref{sec:4C55.16}. We generate four sets of images with different spectral fitting settings:
\begin{itemize}
    \item first-order ordinary polynomial;
    \item second-order ordinary polynomial;
    \item third-order ordinary polynomial;
    \item forced-spectrum fitting, using the smoothed ground-truth spectral index map as input. 
\end{itemize}

All runs produce an image cube, together with a model and a residual image for each of the 80 channels. We extract spectral index maps from these output models: unlike in Sec.~\ref{sec:3C196sim}, we do not smooth the images, because we want to measure the actual spectral index value that each clean component has -- even if it is not fully possible when non-point components are used and hence overlapping occurs, as explained in Sec.~\ref{sec:overlap}. The extraction is performed as usual, fitting each pixel with the power law of Eq.~\eqref{eq:powerlaw} at $\nu_0=150\,\text{MHz}$, and considering only pixels within the $0.40\,\text{mJy\,beam}^{-1}$ contour level of Fig.~\ref{fig:3C196_sim}. While the deconvolved images look similar, the output spectral index maps and residuals are very different, as shown in Fig.~\ref{fig:3C196_rescomp}. It is clear that spectral indices evaluated from ordinary polynomial-fitted models are inaccurate when compared to the ground-truth spectral index map. Looking for instance at hotspot N, we note that Gaussian components in the centre have much steeper spectra than the initial model. Due to the nature of the \textsc{clean} algorithm, such errors cause higher spectral index values for point components in the same region, resulting in an average value that is ultimately consistent with Table~\ref{tab:3C196spix} in such a region. In other words, errors during the deconvolution lead to individual components with extreme spectral indices for which the spatially-integrated flux matches with the data. 

\begin{table}
\caption{Minimum ($\alpha_\text{min}$) and maximum ($\alpha_\text{max}$) values of the spectral index maps obtained from simulated data of 3C\,196 for each of the fitting method and for the non-smoothed ground-truth. Both full pixel distribution and $95^\text{th}$ percentile ranges are provided. Only pixels within the $0.40\,\text{mJy\,beam}^{-1}$ contour level of Fig.~\ref{fig:3C196_sim} are considered.}
\label{tab:3C196spix_out}
\begin{tabular}{lSSSS}
\toprule
\multirow{2}*{Fitting method} & \multicolumn{2}{c}{Full range} & \multicolumn{2}{c}{$95^\text{th}\,\text{percentile}$}\\
\cmidrule(lr){2-3}\cmidrule(lr){4-5}
 & $\alpha_\text{min}$ & $\alpha_\text{max}$ & $\alpha_\text{min}$ & $\alpha_\text{max}$ \\
\midrule
 first-order ordinary pol. & -8.94 & 12.01 & -1.28 & -0.34 \\
 second-order ordinary pol. & -16.05 & 19.69 & -1.36 & -0.39 \\
 third-order ordinary pol. & -25.30 & 44.31 & -1.37 & -0.37 \\
 forced-spectrum & -5.47 & 4.08 & -0.95 & -0.57 \\
 ground-truth & -8.80 & 0.44 & -0.96 & -0.57 \\
\bottomrule
\end{tabular}
\end{table}

\begin{figure*}
    \centering
    \includegraphics[width=1\textwidth]{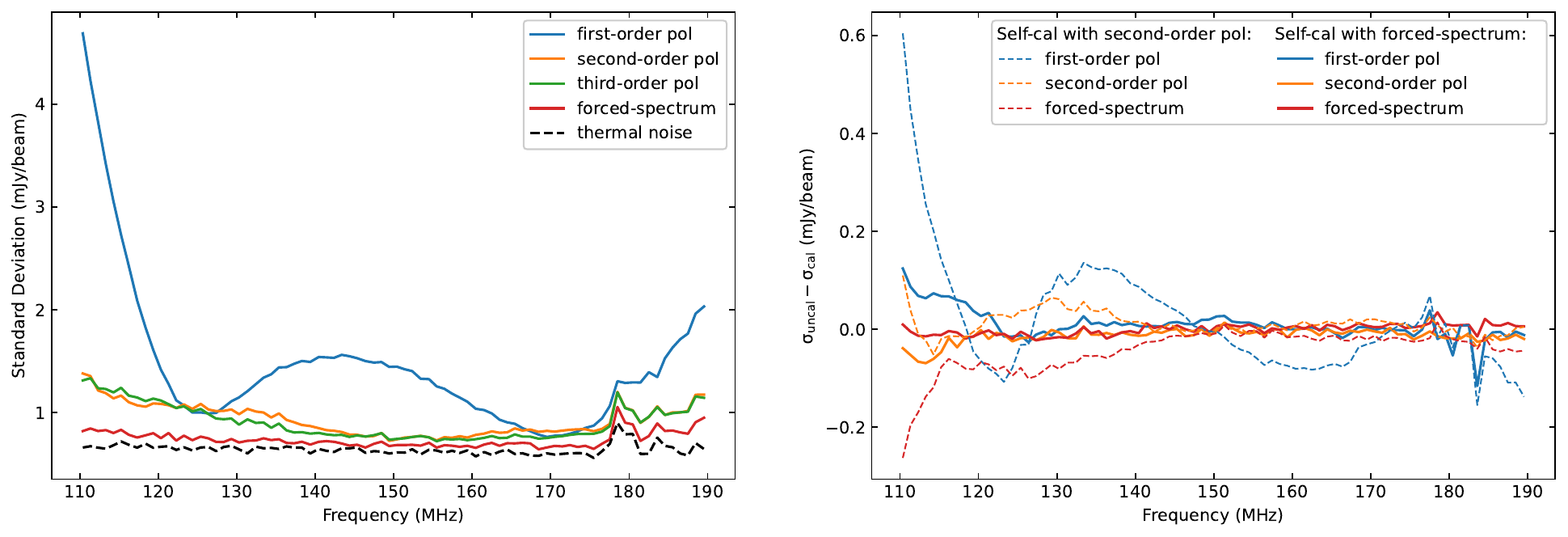}
    \caption{Standard deviation of residuals obtained from deconvolution of initial ($\sigma_\text{uncal}$) simulated 3C\,196 data (left) and its difference with the standard deviation after a self-calibration step ($\sigma_\text{cal}$) using different models (right). Different fitting methods have been used: first (blue), second (orange), and third-order (green) ordinary polynomial functions, and forced-spectrum fitting (red). In the right panel, the self-calibration has been performed with the second-order polynomial (dashed lines) and forced-spectrum (solid lines) models obtained from the deconvolution of the initial data. Image thermal noise is drawn in the left panel with a black dashed line. Only residuals within the dashed contours as Fig.~\ref{fig:3C196_rescomp} are used.}
    \label{fig:3C196_resstat}
\end{figure*}

With the forced-spectrum method, the produced spectral index map is much closer to the ground-truth (top-right panel of Fig.~\ref{fig:3C196_rescomp}) -- even if they do not exactly match when a pixel-based comparison is performed. This is also evident in Fig.~\ref{fig:3C196_spix_distr}, where the pixel distribution of each spectral index map is plotted. The left panel shows the number of pixels with a given spectral index value, binned into bins of 0.01 wide; the right panel shows the pixel-by-pixel difference of ordinary polynomial and forced-spectrum spectral indices, labelled $\alpha$, with the ground-truth ones, labelled $\alpha_\text{GT}$, binned into bins of 0.02 wide. While the left histograms describe the general trend of the output spectral indices with respect to the ground-truth map, the right plot tells us whether $\alpha$ of a given pixel matches $\alpha_\text{GT}$ of the same pixel, giving an approximate estimate of the fitting error. The ground-truth (grey area in left panel) and the forced-spectrum counts (red line in both panels) have been divided by five to make the comparison with ordinary polynomial-fitted values easier, and all the ordinary polynomial fitting histograms have been averaged into a single distribution in the right panel. The gradient due to overlapping components in the extracted ground-truth map is visible in the left panel as single-sided lobes in the grey distribution. Furthermore, when many components overlap, as for lobe W, peaks of the distribution are shifted with respect to the spectral index values of the four components of 3C\,196, marked by the vertical dashed lines. These features occur also for ordinary polynomial and forced-spectrum fitting histograms. Ordinary polynomial fitting of any order generates spectral indices widely spread outside the expected range and with no clearly defined peak. On the other hand, the forced-spectrum distribution follows the ground-truth well.  

The minimum and the maximum spectral indices are provided in Table~\ref{tab:3C196spix_out} for each fitting method. Considering the full range of the distributions, outlier values increase with the order of the ordinary polynomial. Examining the $95^\text{th}$ percentile of the distribution, we notice that forced-spectrum values ($-0.95\le\alpha\le-0.57$) agree with the ground-truth ones ($-0.96\le\alpha\le-0.57$), while ordinary polynomial-fitted spectral indices are outside the expected range. This is confirmed by the right plot of Fig.~\ref{fig:3C196_spix_distr}, where the ordinary polynomial fitting histogram reaches values of $|\alpha-\alpha_{GT}| \approx 0.4$, while, for forced-spectrum, it is confined in the range $|\alpha-\alpha_{GT}| \lesssim 0.1$, with a sharper peak around zero.

Fig.~\ref{fig:3C196_rescomp} also shows residual images obtained from the cleaning at the lower, middle and higher channels, centred at 110.4, 149.4, and 189.5\,MHz, respectively. First-order ordinary polynomial fitting generates the largest residuals due to deconvolution errors, with strong positive emission at the edges of the frequency band and negative emission in the middle. This is not surprising because such kind of fitting models the source spectrum with a straight line, while it is a power law by simulation. The residual standard deviation as a function of frequency is shown in the left panel of Fig.~\ref{fig:3C196_resstat}, where the blue line demonstrate the just described behaviour for the first-order ordinary polynomial function: being high at the edges and in the middle of the band means that the model does not properly reproduce the spectral features of the 3C\,196 simulation. On the other hand, the level of the residuals decreases when higher order ordinary polynomials are used, because curved lines fit a power law more accurately. The standard deviation is extracted from the region where the source has a brightness higher than $0.40\,\text{Jy\,beam}^{-1}$ (superimposed as dashed contour in Fig.~\ref{fig:3C196_rescomp}). In Fig.~\ref{fig:3C196_rescomp}, especially at 110.4\,MHz, we see that second-order ordinary polynomial fitting leaves high residuals in the hotspots and in lobe E, where many point components are present, while a third-order ordinary polynomial more accurately models those components, leaving more uniform residuals. The forced-spectrum method (red line) generates the most uniform and lowest residuals, almost hitting the noise level (black dashed line) at every frequency. Such image noise has been extracted from a Stokes V image and has a frequency-averaged value of $\sigma=0.65\,\text{mJy\,beam}^{-1}$. The standard deviation of the forced-spectrum residuals is on average 1.3 and 1.2 times lower than second and third-order ordinary polynomials, respectively, and 2.0 times lower than first-order ordinary polynomial (up to 5.7 times lower at 110.4\,MHz).

\subsubsection{Self-calibration with second-order ordinary polynomial and forced-spectrum models}

To test the capabilities of the forced-spectrum output model to improve calibration, we compare the difference between self-calibration with and without the forced spectrum method. The self-calibration is performed with settings that are quite common for LOFAR datasets. In particular, we solve for the diagonal gains of the Jones matrices (i.e., the X and Y gain factors) for each channel and time intervals (i.e., one solution every 195\,kHz and 20\,s), and calibrate both phase and amplitude. Without the forced spectrum method, the spectral information used for calibration is obtained during the deconvolution, using second-order ordinary polynomial fitting. The calibrated measurement sets are imaged with the same settings as before, and the forced-spectrum fitting is again performed with the smoothed ground-truth spectral index map of Fig.~\ref{fig:3C196_sim}. 

The results are shown in the right panel of Fig.~\ref{fig:3C196_resstat}, where the standard deviation of the residuals, labelled $\sigma_\text{cal}$, is subtracted frequency-by-frequency from the uncalibrated one, labelled $\sigma_\text{uncal}$, plotted in the left panel. Whereas $\sigma_\text{cal}>\sigma_\text{uncal}$ means that the calibration degrades the image quality, $\sigma_\text{cal}<\sigma_\text{uncal}$ does not necessarily mean the opposite, because our already-perfectly simulated data can not be improved by calibration. When this happens, it is likely that the true sky signal or noise is absorbed into the calibration gains.

The second-order ordinary polynomial calibrated data show an overall degradation, especially at low frequencies with first-order ordinary polynomial fitting. Forced-spectrum residuals have $\sigma_\text{cal}>\sigma_\text{uncal}$ because we are fitting a power law to data that follow a parabolic curve after the self-calibration step. The line for second-order ordinary polynomials oscillates around $\sigma_\text{uncal}-\sigma_\text{cal}=0$: forcing spectral behaviour that is not intrinsically in the data increases deconvolution errors, leaving such spectral features in the residuals. This demonstrates the need to calibrate with the best possible model, which must be both spatially and spectrally accurate. Using the forced-spectrum model for the calibration leaves residuals almost unaffected, even if some oscillations are visible because some signal absorption into gains occurs during the calibration. 

\subsection{Results from 4C+55.16 observation}\label{sec:4C55.16res}

We test the forced-spectrum method on real observed data to understand how it works when systematics, such as calibration errors, are present. We chose 4C+55.16 as test case, whose self-calibration and spectral index extraction have been described in Sec.~\ref{sec:4C55.16}. We make images following the same procedure used for the 3C\,196 simulation in the previous section, but split the bandwidth into 92 channels as we did for the latest self-calibration steps of 4C+55.16. We make a set of images with the first, second, and third-order ordinary polynomial fits, that we compare to the forced-spectrum method using the spectral index map of Fig.~\ref{fig:4C55_image}. 

In this section we will discuss two different results from 4C+55.16: results before and after an extra self-calibration step. To distinguish between the earlier self-calibration steps that were performed, we will refer to this step as the ``final'' self-calibration step. The final self-calibration step is performed using the model obtained from the forced-spectrum deconvolution. This will therefore also test whether the use of a forced-spectrum model may improve calibration.

\subsubsection{Before final self-calibration}

\begin{figure*}
    \centering
    \includegraphics[width=1\textwidth]{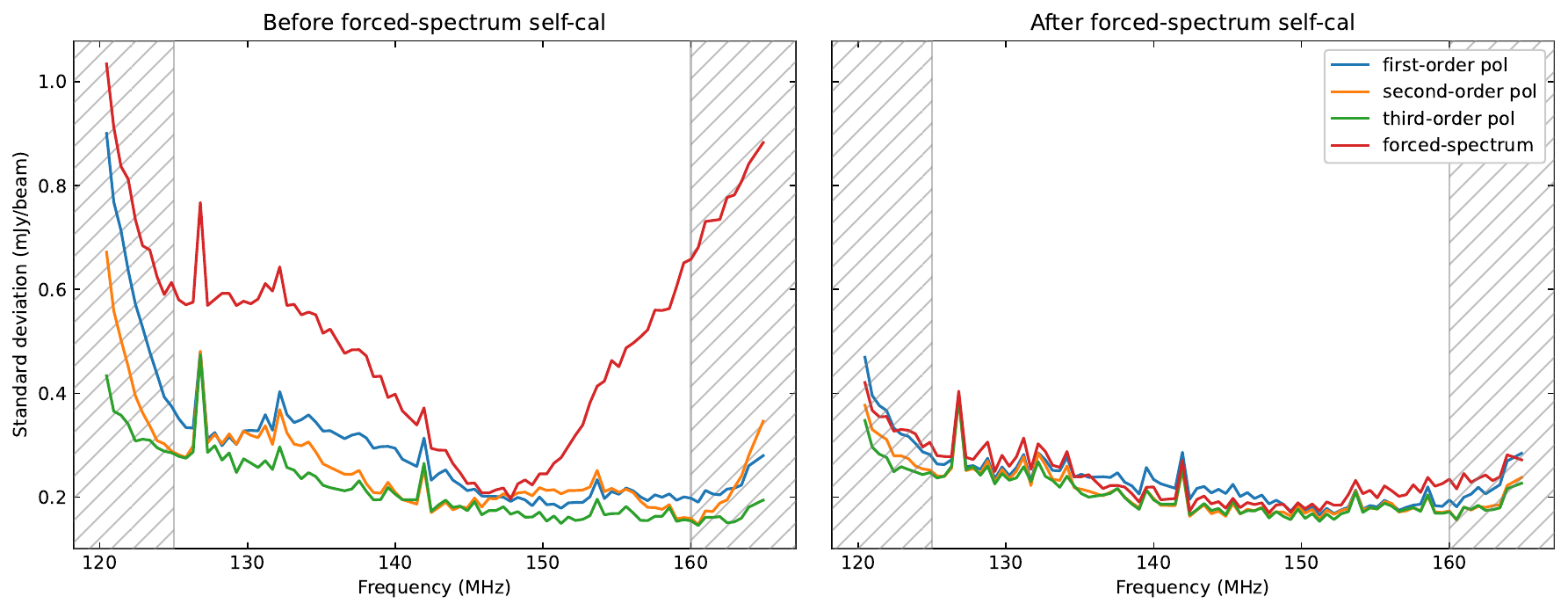}
    \caption{Standard deviation of smoothed residuals obtained from deconvolution of 4C+55.16 data before (left) and after (right) the last self-calibration using the forced-spectrum model. First (blue), second (orange), and third-order (green) ordinary polynomial functions, and forced-spectrum fitting (red) are shown. The hatched areas mark the frequencies excluded from the analysis (we consider the range 125--160\,MHz). Only residuals within the $20\sigma$ contour level of Fig.~\ref{fig:4C55_image} are used.}
    \label{fig:4C55_resstat}
\end{figure*}

We will start by discussing the results before the final self-calibration. As explained in Sec.~\ref{sec:overlap}, in the model catalogue (an example from 4C+55.16 is provided in Appendix~\ref{app:modeltext}), each component has the expected spectral index from the input map based on its coordinates, because components are listed with their central position and spectral index in the model catalogue. However, the residuals from the forced-spectrum method are higher than those from the ordinary polynomial fits. The standard deviation of the 92 residual images are shown in the left panel of Fig.~\ref{fig:4C55_resstat} for the four fitting methods, extracted from the $20\sigma$ contour level of Fig.~\ref{fig:4C55_image} after smoothing using a circular Gaussian kernel with a FWHM of 500\,mas.\footnote{We removed the output channels centred at 136.08, 140.97, and 164.40\,MHz because of bad image quality due to interference at these frequencies.} This reduces the contribution of high resolution errors that arise because the cores are inaccurately modelled; since these are caused by the fundamental resolving resolution of the instrument and not by the spectral modelling, we do not want our residual images to be dominated by those errors. Residuals are higher at the band edges because fitting is less accurate here. For this reason, we will focus our analysis only on the range 125--160\,MHz. Third-order ordinary polynomial fitting results in the lowest residuals. This may be caused by the previous self-calibration steps, where we used models obtained by fitting ordinary polynomial functions of such order. However, the other ordinary polynomial results show comparable residuals, especially at frequencies higher than $146\,\text{MHz}$, in contrast to what we have seen in the simulations (see Fig.~\ref{fig:3C196_resstat}). First-order ordinary polynomial fitting has the largest residuals, which are on average 1.3 times higher than other ordinary polynomial fitting methods, up to 1.5 times higher at 139.0\,MHz. Such small differences highlight how higher-order ordinary polynomial fits do not significantly improve residuals when systematics are present in the data \citep{offringa_etal:2016}. 

Unlike in our simulations, the forced-spectrum method produces higher residuals. They are on average 2.1 times higher than third-order ordinary polynomial fitting, with differences getting larger toward both ends of the band, and is 4.2 times higher at 159.5\,MHz. Most of such excess residual power comes from the core region, which switches from positive to negative residual brightness from low to high frequency. The (smoothed) maximum residual brightness is $6.6\,\text{mJy\,beam}^{-1}$ at 125.3\,MHz, while the minimum is $-11.1\,\text{mJy\,beam}^{-1}$ at 159.5\,MHz. Since $\alpha>0$ for the core, this means that the spectral index estimated for the core is too high. In the range 145--150\,MHz the forced-spectrum is comparable to ordinary polynomial fitting. Furthermore, the excess residual power becomes stronger at higher frequencies. Part of this might be explained by the intrinsic double nature of the core, which we modelled as a single component with a single spectral index value, while we know that the two components have different spectral signatures. 

\subsubsection{After final self-calibration}

\begin{figure*}
    \centering
    \includegraphics[width=1\textwidth]{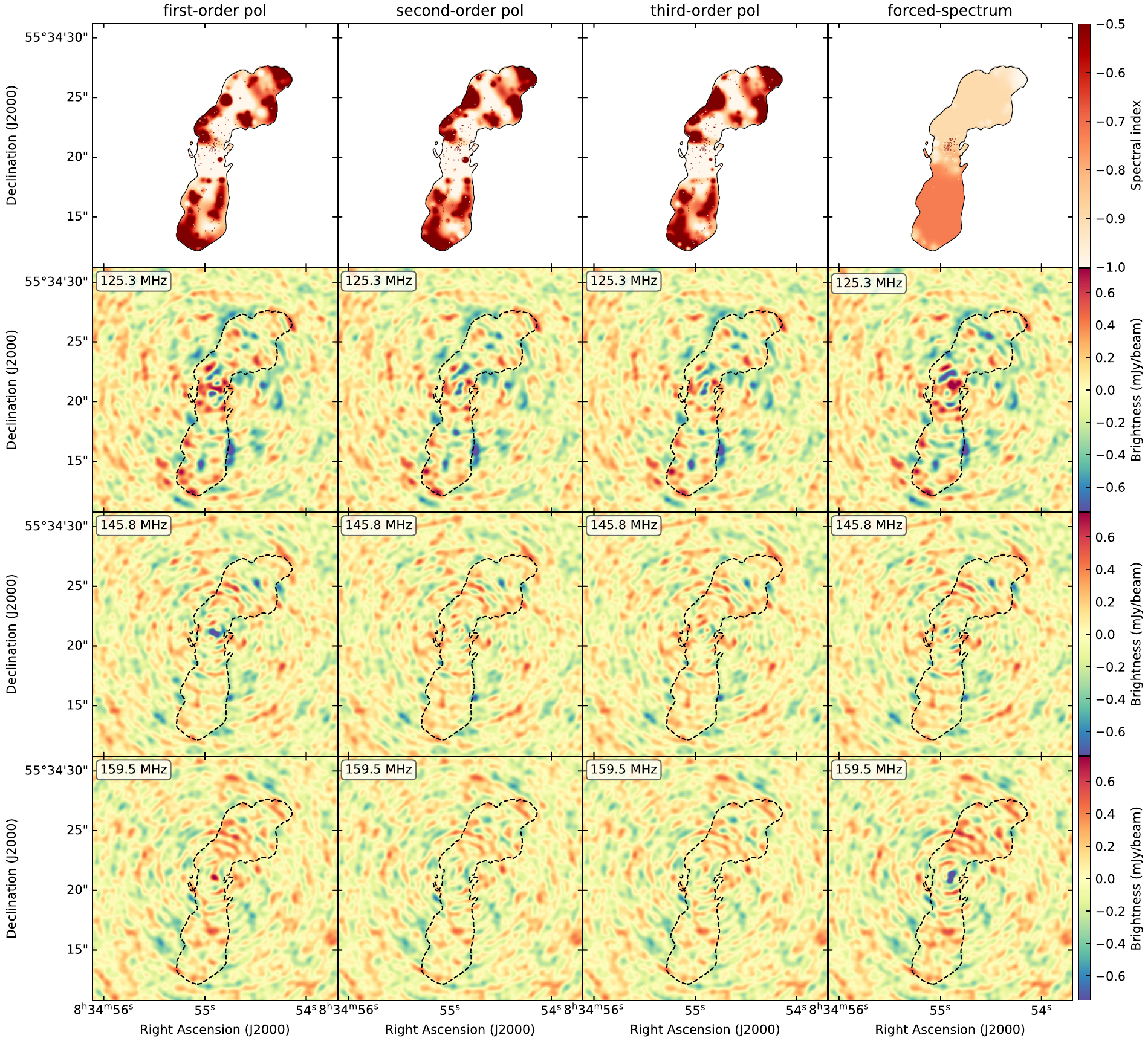}
    \caption{Deconvolution results of 4C+55.16 data with different fitting methods after final self-calibration using forced-spectrum output models. From left to right: results from first, second, and third-order ordinary polynomial fitting, and from the forced-spectrum fitting. From top to bottom: spectral index maps from output models, and residual images of a low, middle, and high channels, centred at 125.3, 145.8, and 159.5\,MHz, respectively, avoiding the edges of the observed band. Residuals images have been smoothed by a circular Gaussian kernel with a FWHM of 500\,mas to enhance the strong features. The colour scale of spectral index maps is the same of Fig.~\ref{fig:4C55_image} to make the comparison easier with the forced-spectrum input map; however, values from ordinary polynomial fits can be outside that range, as shown in Fig.~\ref{fig:4C55_spix_distr} and in Table~\ref{tab:4C55.16spix_out}. On the other hand, spectral indices obtained from the forced-spectrum method are similar to the expected values with only few outlier pixels. The dashed contours superimposed on the residual images indicate the $20\sigma$ level of Fig.~\ref{fig:4C55_image}.}
    \label{fig:4C55_rescomp}
\end{figure*}

We will now discuss the results after the final self-calibration. This step is performed using the forced-spectrum output model, solving for the diagonal gains of the Jones matrices for each channel and time intervals (i.e., one solution every 0.5\,MHz and 16\,s), and calibrating both phase and amplitude. We make new sets of images, still using the initial input map for the forced-spectrum fitting.

The final results are summarised in Fig.~\ref{fig:4C55_rescomp}. The top row shows the spectral index maps directly extracted (i.e., with no smoothing) from the output models for the four fitting methods, considering only pixels within the $20\sigma$ contour level of Fig.~\ref{fig:4C55_image}, this time matching what we have found from the simulations: forced-spectrum method generates more spectrally accurate models than ordinary polynomial fits. Ordinary polynomial-fitted models deviate from the expected values especially at the edges of the source, where the signal-to-noise is lower. This happens because images in those regions are dominated by calibration and deconvolution artefacts, which can be picked up as clean components during deconvolution. The spectral index of the core is a combination of point components and Gaussians. For some of these Gaussians, their centres fall into the lobes N and S. The extracted spectral index from this underlying diffuse emission is therefore an average between the values of the lobe N and lobe S for the forced-spectrum output models (see Sec.~\ref{sec:overlap}).

\begin{figure*}
    \centering
    \includegraphics[width=1\textwidth]{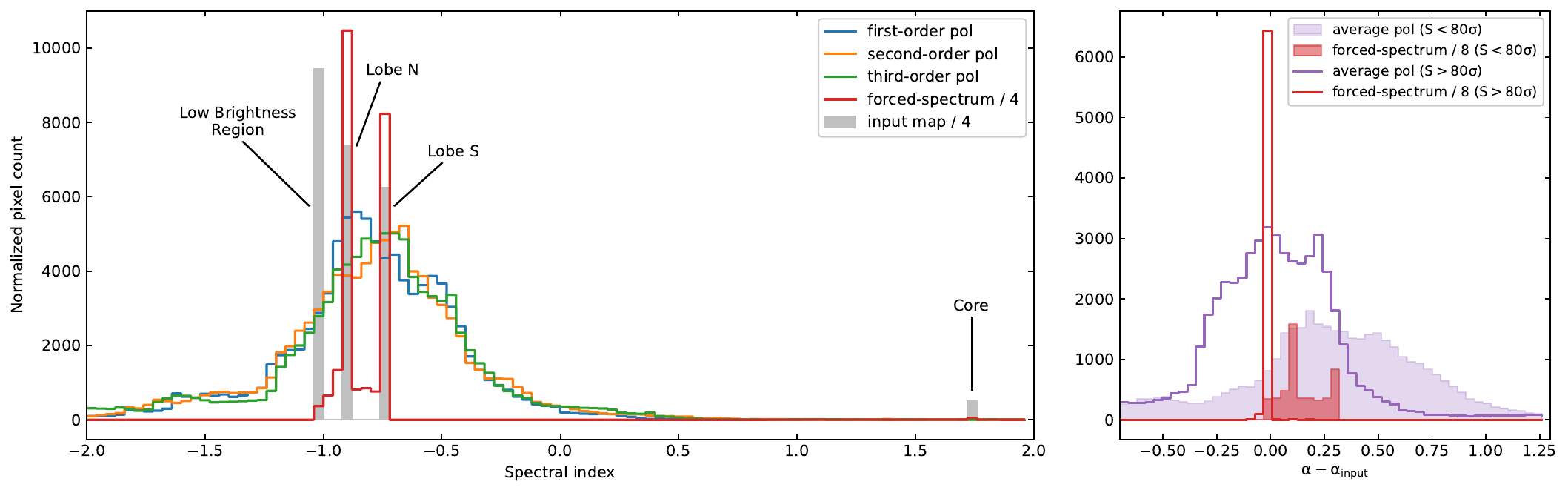}
    \caption{Pixel distributions of the spectral index maps shown in Fig.~\ref{fig:4C55_rescomp} (left) and of their difference with the input ($\alpha_\text{input}$) map (right). First (blue), second (orange), and third-order (green) ordinary polynomial fitting distributions are plotted in comparison with the forced-spectrum (red) and the input map (grey area) ones. In the right panel all the polynomial fitting histogram have been averaged into a single distribution (purple). Here, empty and filled histograms respectively show distributions of pixels with flux density higher and lower than the $80\sigma$ level of Fig.~\ref{fig:4C55_image}. The histograms are obtained by binning spectral indices into bins of 0.04 wide in both panels. Forced-spectrum and input map values have been divided by 4 in the left panel and by 8 in the right panel, to make the comparison with polynomial distributions easier.}
    \label{fig:4C55_spix_distr}
\end{figure*}

\begin{table}
\caption{Minimum ($\alpha_\text{min}$) and maximum ($\alpha_\text{max}$) values  of the spectral index maps obtained from the final self-calibrated data of 4C+55.16 for each of the fitting methods. Both full pixel distribution and $95^\text{th}$ percentile ranges are provided. Only pixels within the $20\sigma$ contour level of Fig.~\ref{fig:4C55_image} are considered.}
\label{tab:4C55.16spix_out}
\begin{tabular}{lSSSS}
\toprule
\multirow{2}*{Fitting method} & \multicolumn{2}{c}{Full range} & \multicolumn{2}{c}{$95^\text{th}\,\text{percentile}$}\\
\cmidrule(lr){2-3}\cmidrule(lr){4-5}
 & $\alpha_\text{min}$ & $\alpha_\text{max}$ & $\alpha_\text{min}$ &  $\alpha_\text{max}$ \\
\midrule
 first-order ordinary pol. & -12.27  & 24.64 & -1.82 & -0.11 \\
 second-order ordinary pol. & -24.26 & 49.38 & -1.70 & -0.31 \\
 third-order ordinary pol. & -29.48 & 54.34 & -1.96 & 0.21 \\
 forced-spectrum & -1.20 & 2.92 & -0.99 & -0.72 \\
 input map & -1.02 & 1.74 & -1.02 & -0.72 \\
\bottomrule
\end{tabular}
\end{table}

Pixel distributions of the spectral index maps are shown in Fig.~\ref{fig:4C55_spix_distr}. The left panel shows the number of pixels with a given spectral index value, whereas the right panel shows the pixel-by-pixel difference of ordinary polynomial and forced-spectrum spectral indices, labelled $\alpha$, with the input ones, labelled $\alpha_\text{input}$, both binned into bins of 0.04 wide. Input map (grey area in left panel) and forced-spectrum counts (red histograms in both panels) have been divided by 4 in the left panel and by 8 in the right panel, where all the ordinary polynomial fitting histograms have been averaged into a single distribution (purple histograms). Furthermore, the right panel shows two kind of distributions for each fitting method: empty histograms represent pixels with $S(\nu_0)>80\sigma$ (i.e., core and lobes regions of Fig.~\ref{fig:4C55_image}), while full histograms represent pixels with $20\sigma < S(\nu_0) < 80\sigma$ (i.e., surrounding low brightness emission), where $\sigma=80\,\text{\textmu Jy\,beam}^{-1}$. The forced-spectrum method generates spectral indices that agree well with the input map, while ordinary polynomial fits generate values with a larger spread, with $|\alpha-\alpha_\text{input}|\approx0.7$ for the core and lobes regions, reaching differences higher than 1 for the low brightness  emission. Ordinary polynomial fits tend to generate $\alpha > \alpha_\text{input}$ in low brightness regions, while $\alpha<\alpha_\text{input}$ in high brightness ones. This is also visible in Table~\ref{tab:4C55.16spix_out}. 

Modelling the core with only a few point components causes a smaller peak at the input spectral index value, since most of the core pixels contains the diffuse emission described above. This generates small gradients in the forced-spectrum histogram between the three main peaks. However, such gradients are mainly due to the leaking of the Gaussian components of the lobes into the surrounding low brightness region. Many pixels from this region take the spectral index values of lobe N and lobes S, resulting in a number of pixels with those values higher than the input ones. The right panel clearly shows this effect: for fluxes lower than $80\sigma$, two peaks arise at $\alpha-\alpha_\text{input}=0.12$ and 0.30. Since $\alpha_\text{input}=-1.02$ in this region, we recover $\alpha=-0.90$ and $-0.72$, which are the input values for lobe N and lobes S, respectively. 

Residual images are also shown in Fig.~\ref{fig:4C55_rescomp}, obtained from a low, middle, and high channels, centred at 125.3, 145.8, and 159.5\,MHz, respectively. As before the final self-calibration, we smooth the images with a circular Gaussian filter with a FWHM of 500\,mas. Looking at the residuals within the dashed contour line, representing the $20\sigma$ level, it is evident that differences between the four fitting methods are negligible in the lobe regions, but not in the core. Residuals from the forced-spectrum fitting still show an excess brightness that is positive at low frequency and negative at high frequency, becoming comparable to second and third-order ordinary polynomial fittings in the middle of the band. However, the self-calibration has reduced the maximum and minimum residual brightness at 125.3 and 159.5\,MHz by a factor of ${\sim}6$, which now are 1.2 and $-1.9\,\text{Jy\,beam}^{-1}$, respectively. Extracting the standard deviation as we did for the left panel of Fig.~\ref{fig:4C55_resstat}, we see that the differences are almost negligible between all the fitting methods, as shown in the right panel of Fig.~\ref{fig:4C55_resstat}. Forced-spectrum residuals are on average only 1.1 times higher than third-order ordinary polynomial residuals, which are the lowest over most of the band. The strongest difference is toward the high-end of the band, and is of the order of 30\%. As expected, first-order ordinary polynomial fitting generates the highest residuals in the middle of the band, because a straight line fails to fit a power law. However, even in this case, the difference is just 24\% with the third-order ordinary polynomial. In addition, the overall standard deviation is lower than before the final self-calibration. This means that self-calibrating with the models obtained from the forced-spectrum fitting improves the overall data quality, especially at the low-end of the band.

\section{Discussion and conclusions}\label{sec:conclusions}

In this paper, we have presented a novel method -- implemented in \textsc{wsclean} -- for generating sky models with physical spectral information, by taking advantage of previously created spectral index maps. While the method itself does not generate spectral index information, it allows for the decomposition of convolved data into components. This produces a model that closely matches the data and can be used for further calibration or source subtraction. The input spectral index map is used inside the multi-frequency (MF) and multi-scale (MS) \textsc{clean} algorithm, and constrains the spectral index term of every found component. As a consequence, only a single scaling factor must be calculated through a modified-weighted linear least-squares fit (Sec.~\ref{sec:method_fs}). Thus, the forced-spectrum method reduces inaccuracies that are common in typical MF fitting methods, particularly for observations with small bandwidth, and generates accurate spectral models directly during the deconvolution. Any type of spectral index map can be used, whether from multi-instrument observations or in-band data, providing flexibility in how the spectral information that is transferred into the sky model is determined. Because the forced-spectrum method assigns the spectral index to each cleaned component based on its central position, inaccuracies in the pixel-based spectral index maps may result when extended (i.e., Gaussian) components overlap (Sec.~\ref{sec:overlap}). This spectral index mixing primarily affects the low brightness parts of a source and generates gradients between regions with different spectral indices. The overlapping issue can be avoided using \textsc{clean} algorithms that are only based on point components, but the benefits of using the MS deconvolution outweigh this minor issue.

We also demonstrate a clustering method for extracting spectral indices from in-band observations (Sec.~\ref{sec:4C55.16}). The use of MF deconvolution in such observations often generates output channel images that are dominated by calibration and deconvolution errors, making it challenging to determine accurate spectral indices. To overcome this, in the clustering method we divide the source into a certain number of regions, calculate the weighted average of the brightness within each region, and extract the spectral indices. This reduces the effect of systematics.

Spectrally-forced imaging provides a new tool that can be used to improve generic MF imaging. We have focused our testing on the imaging of individual, non-varying sources. We have not yet tested wide-field imaging or deconvolution of transients and point sources. The method may improve results for those use-cases as well, but it is likely that its strength lies in the modelling of individual, persistent, and resolved sources that are relatively challenging to model.

The forced-spectrum method works well in combination with the MF-MS deconvolution of \textsc{wsclean} when a model catalogue is also generated. Since such a model is not limited by the pixel scale, it can be used to save high-resolution models of calibrators and strong sources that could be used by different experiments, reducing the required data volume. Providing well-motivated external spectral information into such models is crucial for high-resolution and wide-field interferometers operating at low-frequencies, such as LOFAR and the upcoming SKA-Low\footnote{Square Kilometers Array, \url{https://www.skao.int/en/explore/telescopes}} \citep{dewdney_etal:2009,braun_etal:2019}, which have the capability to generate images ${\sim}100\,000$ pixels in size.

Results from both simulations (Sec.~\ref{sec:3C196sim_res}) and real observations (Sec.~\ref{sec:4C55.16res}) have shown an improvement with respect to typical ordinary polynomial fitting methods, which generate inaccurate spectral indices, especially for faint sources. Promising results are also observed in the residual images, which appear more uniform and lower in magnitude in the simulations. On the other hand, real observations are dominated by other systematics, such as calibration and deconvolution errors. These can be over-fitted and absorbed by high-order ordinary polynomial fits -- into incorrect spectral indices -- and eventually produce better residuals than forced-spectrum fitting, which does not allow the same freedom and such systematic spectral features are visible in the residual images. Despite this, we have shown that more accurate models are obtained and that they can be used for calibration, ultimately improving the overall data quality, resulting in residuals that are similar to those obtained with ordinary polynomial fits. The models generated by forced spectral fitting can also be more easily extrapolated to different frequencies. This may prove to be useful for sharing models between instruments.

An interesting other application of the forced-spectrum method could be to determine how well a spectral index map matches the data, by looking at the residuals after forced-spectrum deconvolution. Image residuals may also be used to adjust the input spectral index map, especially if it has been made with the clustering method. For example, if the residual brightness of pixels or regions go from positive (negative) at low frequency to negative (positive) at high frequency, the spectral indices of those components should be steeper (flatter) than the input ones. The adjusted spectral index map better matches the data and can be used to output a model that can then be used to self-calibrate the dataset.

These results are highly relevant for improving upper limits on the 21-cm power spectrum from the Epoch of Reionization and Cosmic Dawn. With calibration errors below $0.1\%$ being critical for achieving the high dynamic range needed to measure the 21-cm power spectrum during these cosmic eras \citep{mazumder_etal:2022}, the sky models derived from forced-spectrum fitting hold significant promise for enhancing the accuracy of such measurements.

\section*{Acknowledgements}
EC, ARO and LVEK would like to acknowledge support from the Centre for Data Science and Systems Complexity (DSSC), Faculty of Science and Engineering at the University of Groningen. LVEK and BKG acknowledge the financial support from the European Research Council (ERC) under the European Union's Horizon 2020 research and innovation programme (Grant agreement No.\ 884760, ``CoDEX''). FGM acknowledges support from a PSL Fellowship. RT and RJvW acknowledge support from the ERC Starting Grant No.\ 804208, ``ClusterWeb''. LOFAR, the Low Frequency Array designed and constructed by ASTRON, has facilities in several countries, that are owned by various parties (each with their own funding sources), and that are collectively operated by the International LOFAR Telescope (ILT) foundation under a joint scientific policy.

In this work, we made use of the \textsc{kvis} \citep{gooch:1996} and \dsnine\ \citep{joye_mandel:2003} FITS file image viewers, and the \textsc{astropy} \citep{astropy_coll:2022}, \textsc{matplotlib} \citep{hunter:2007}, \textsc{numpy} \citep{harris_etal:2020}, \textsc{pandas} \citep{mckinney:2010}, \textsc{scipy} \citep{virtanen_etal:2020} \textsc{python} packages. 

\section*{Data Availability}
The data underlying this article will be shared on reasonable request to the corresponding author.

\bibliographystyle{mnras}
\input{MAIN_spix_submitted.bbl}



\appendix

\section{WSClean model catalogue}\label{app:modeltext}
In Table~\ref{tab:4C55.16comp_list} we report four example components from the clean component list file that \textsc{wsclean} produces, one line for each of the regions into which we have divided 4C+55.16 (see Sec.~\ref{sec:4C55.16}). More details about this file format can be found at \url{https://wsclean.readthedocs.io/en/latest/component_list.html}.

\begin{table*}
\caption{Four lines from the clean component list file after the first forced-spectrum fitting. From top to bottom: point component from the core region, and Gaussian components from lobe S, lobe N, and from the surrounding regions. The original file provided also the reference frequency at which the flux density $I$ (i.e., $S$ with the notation adopted in this paper) is estimated, which is 144.30\,MHz for all the components, and the orientation of the Gaussians, which is always zero because only circular Gaussians are modelled by the MS deconvolution.}
\label{tab:4C55.16comp_list}
\begin{tabular}{lccccSccc}
\toprule
Name & Type & RA & Dec & $I$ & $\text{SpectralIndex}$ & MajorAxis & MinorAxis & Orientation \\
 & & & & (Jy) & & (arcsec) & (arcsec) & (deg) \\
\midrule
s0c1069 & POINT & $08^\text{h}34^\text{m}54\rlap{.}^\text{s}906$ & $55^\circ34'20\rlap{.}''950$ & $0.293$ & 1.738 & -- & -- & -- \\
s3c41 & GAUSSIAN & $08^\text{h}34^\text{m}54\rlap{.}^\text{s}971$ & $55^\circ34'18\rlap{.}''750$ & $0.010$ & -0.724 & $1.373$ & $1.373$ & 0 \\
s4c30 & GAUSSIAN & $08^\text{h}34^\text{m}54\rlap{.}^\text{s}567$  & $55^\circ34'24\rlap{.}''950$ & $0.149$ & -0.897 & $2.746$ & $2.746$ & 0 \\
s1c0 & GAUSSIAN & $08^\text{h}34^\text{m}55\rlap{.}^\text{s}080$ & $55^\circ34'12\rlap{.}''375$ & $0.002$ & -1.018 & $0.343$ & $0.343$ & 0 \\
\bottomrule
\end{tabular}
\end{table*}

\section{Spectral indices observed in pixel-based models}\label{app:overlapping}

Here, we derive the spectral index that is observed in a model image when components overlap. The situation is described in Fig.~\ref{fig:si_overlapping}. 
We consider a Gaussian component with the peak in position A and a point component in position B, whose spectra are described by a power law, with spectral indices $\alpha_\text{A}$ and $\alpha_\text{B}$, respectively, constant over their full shape. The flux density of the Gaussian is
\begin{equation}\label{eq:SG}
    S_\text{G}(\nu) = S_\text{G}(\nu_0) \left(\frac{\nu}{\nu_0}\right)^{\alpha_\text{A}} \, ,
\end{equation}
where $\nu_0$ is the reference frequency at which the normalisation is evaluated, and similarly for the point source $S_\text{P}$, with spectral index $\alpha_\text{B}$. Since $\alpha_\text{A}\ne\alpha_\text{B}$, there is a frequency $\nu_\text{int}$ where the two spectra intersect; without loss of generality, we can choose to normalise the flux density at $\nu_0=\nu_\text{int}$, so that $S_\text{G}(\nu_0)=S_\text{P}(\nu_0)$ in B, where the components coincide.

When the spectral index of the resulting (gridded) model image is observed, the resulting flux density in B is the sum of $S_\text{G}(\nu)$ and $S_\text{P}(\nu)$, such that
\begin{equation}
    S(\nu) = S_\text{P}(\nu_0)\left[\left(\frac{\nu}{\nu_0}\right)^{\alpha_\text{A}} + \left(\frac{\nu}{\nu_0}\right)^{\alpha_\text{B}}\right],
\end{equation}
which is not a power law, as shown in the bottom panel of Fig.~\ref{fig:si_overlapping}. However, when we analyse the produced spectral indices (as in the first row of Fig.~\ref{fig:3C196_rescomp}), we fit a power law pixel-by-pixel: in this case, the resulting spectral index $\alpha$ is a combination of $\alpha_\text{A}$ and $\alpha_\text{B}$ and changes with the frequency range over which the fit is performed. 

To better understand what is the expected $\alpha$ in B, we move to a log-log space, so that the flux density is now defined as
\begin{equation}\label{eq:logS_sum}
    \log [S(\nu)] = \log [S_\text{P}(\nu_0)] + \log\left[\left(\frac{\nu}{\nu_0}\right)^{\alpha_\text{A}} + \left(\frac{\nu}{\nu_0}\right)^{\alpha_\text{B}}\right].
\end{equation}
Then, the spectral index -- or the slope -- of $S(\nu)$ can be calculated by the derivative of $\log S$ with respect to $\log\nu$:
\begin{equation}\label{eq:dlogS_dlognu}
    \alpha(\nu)=\frac{\text{d}\log S}{\text{d} \log \nu} = \frac{\alpha_\text{A}\nu_0^{\alpha_\text{B}}\nu^{\alpha_\text{A}} + \alpha_\text{B}\nu_0^{\alpha_\text{A}}\nu^{\alpha_\text{B}} }{\nu_0^{\alpha_\text{B}}\nu^{\alpha_\text{A}} + \nu_0^{\alpha_\text{A}}\nu^{\alpha_\text{B}}}\, ,
\end{equation}
which is a function of $\nu$, as expected. 

With $\alpha_\text{A}>\alpha_\text{B}$, ce can consider three special cases: (1) if $\nu \ll \nu_0$, we can re'write Eq.~\eqref{eq:dlogS_dlognu} as
\begin{equation}
    \alpha(\nu) = \frac{\alpha_\text{A}\left(\nu/\nu_0\right)^{\alpha_\text{A}-\alpha_\text{B}} + \alpha_\text{B}}{\left(\nu/\nu_0\right)^{\alpha_\text{A} -\alpha_\text{B}} + 1}\, ,
\end{equation}
from which we find $\alpha \approx \alpha_\text{B}$; (2) if $\nu \gg \nu_0$, we can re'write Eq.~\eqref{eq:dlogS_dlognu} as
\begin{equation}
    \alpha(\nu) = \frac{\alpha_\text{A} + \alpha_\text{B}\left(\nu/\nu_0\right)^{\alpha_\text{B}-\alpha_\text{A}}}{\left(\nu/\nu_0\right)^{\alpha_\text{B} -\alpha_\text{A}} + 1}\, ,
\end{equation}
from which we find $\alpha \approx \alpha_\text{A}$; (3) if $\nu\approx\nu_0$, we obtain
\begin{equation}
    \alpha \approx \frac{\alpha_\text{A} + \alpha_\text{B}}{2}\,. 
\end{equation}
This demonstrates that the slope of the sum of two power laws is $\alpha_\text{B}\lesssim \alpha(\nu)\lesssim \alpha_\text{A}$. 

\section{Self-calibration of 4C+55.16 data}\label{app:self-cal_4c55}

Here, we provide a more detailed description of the self-calibration process that was performed to improve the quality of the 4C+55.16 data, in order to meet the requirements for testing the forced-spectrum method. The models used for this operation are obtained from imaging the data with \textsc{wsclean}. 

We start splitting the full bandwidth of 45\,MHz into 12 output channels, each ${\sim}4\,\text{MHz}$ wide, which are jointly cleaned with no MF weighting. Using uniform weighting results in an integrated synthesised beam with a FWHM of $207\times143\,\text{mas}$. The smallest synthesised beam is achieved for the image in the frequency range 161--165MHz, with a FWHM of $128\times184\,\text{mas}$. To produce shallow Stokes I images, we use a pixel scale of $25\times25\,\text{mas}$ and the W-gridding algorithm, which is more accurate and -- in our case -- faster than the standard W-stacking method of \textsc{wsclean} \citep{arras_etal:2021,Ye_etal:2022}. We clean down to an initial threshold of $50\sigma$ and a final threshold of $10\sigma$, where $\sigma=90\,\text{\textmu Jy\,beam}^{-1}$ for the frequency-integrated image. The auto-masking algorithm was used to generate scale-dependent masks that were used to constrain the cleaning. Using such high thresholds ensures that no artefacts are included in the models and that visibilities are calibrated by real components only. The obtained model is directly saved in the measurement set file of the data. 

The calibration step is then performed with \dppp. We use the output models within the \texttt{gaincal} algorithm to solve only the diagonal phases of the Jones matrix (i.e., a $2\times2$ complex matrix for every station) for each channel and 4 time intervals (i.e., one solution every 64\,s). These solutions are then applied to the data and a new imaging step is started. 

In the next self-calibration iterations, we gradually decrease the imaging thresholds, reaching $12\sigma$ and $3\sigma$ for the initial and the final thresholds. This is possible because the image quality improves after each iteration, going from an initial image dynamic range of $23\,000$ to $30\,000$ after the last iteration, where we obtain the deepest integrated image, with a noise of $67\,\text{\textmu Jy\,beam}^{-1}$. Both the image noise and the dynamic range are always evaluated from imaging 12 output channels with an initial threshold of $8\sigma$ and a final threshold of $3\sigma$ through all the self-calibration step, to make the comparison as fair as possible. In addition to decreasing the cleaning thresholds, also the solution interval is reduced from 4 time steps to 2 during the second self-calibration iteration, and then to 1 (i.e., one solution every 16\,s) from the third iteration to the end. However, solving with a model every 4\,MHz does not completely solve the phase problem that causes artefacts around the core, because a single solution is applied to all the channels within that frequency range. In fact, issues are still present when every frequency channel is imaged. For this reason, at a certain iteration of the self-calibration we start imaging every channel individually, obtaining 92 output images. Using 92 channels with no MF weighting degrades the synthesised beam: the integrated image now has a beam with a FWHM of $255\times184\,\text{mas}$, while the best synthesised beam is achieved in the range $163-163.5\,\text{MHz}$ with a FWHM of $166\times225\,\text{mas}$. 

After having solved the calibration issue in the core structure, we perform a couple more iterations solving diagonal gains (i.e., both amplitude and phase in the diagonal of the Jones matrix), to make sure that the spatially-integrated spectrum of the source follows the expected slope. To do this, we re-scale the output models to a flux density of 8.66\,Jy at 142.67\,MHz using a power law with $\alpha=-0.02$ (see Sec.~\ref{sec:4C55.16}). This model is then predicted into the measurement set file and used to evaluate the gains.


\bsp	
\label{lastpage}
\end{document}